\documentclass[a4paper,fleqn,usenatbib]{mnras}

\usepackage{newtxtext,newtxmath}
 
\usepackage[T1]{fontenc}
\usepackage{ae,aecompl}

\usepackage{graphicx}	
\usepackage{amsmath}	
\usepackage{amssymb}	
\usepackage{mathtools}
\usepackage[usenames,dvipsnames]{xcolor}

\newcommand{\ie}{i.e.,~}
\newcommand{\eg}{e.g.,~}

\title[Comparison of the temperature ratio
  prescription]{Comparison of the ion-to-electron temperature ratio
  prescription: GRMHD simulations with electron thermodynamics}

\author[Y. Mizuno et al.]
{Yosuke Mizuno$^{1,2}$\thanks{E-mail: mizuno@sjtu.edu.cn (YM)},
Christian M. Fromm$^{3,2,4}$,
Ziri Younsi$^{5,2}$,  
Oliver Porth$^{6}$, \newauthor
Hector Olivares$^{7,2}$, 
Luciano Rezzolla$^{2,8,9}$
\\
$^{1}$Tsung-Dao Lee Institute and School of Physics \& Astronomy, Shanghai Jiao-Tong University, Shanghai, 200240, People's Republic of China\\
$^{2}$Institut f\"{u}r Theoretische Physik, Goethe Universit\"{a}t, Max-von-Laue-Str. 1, 60438 Frankfurt am Main, Germany\\
$^{3}$Black Hole Initiative at Harvard University, 20 Garden Street, Cambridge, MA 02138, USA\\
$^{4}$Max-Planck-Institut f\"{u}r Radioastronomie, Auf dem H\"{u}gel 69, D-53121 Bonn, Germany\\
$^{5}$Mullard Space Science Laboratory, University College London, Holmbury St. Mary, Dorking, Surrey RH5 6NT, UK\\
$^{6}$Anton Pannekoek Institute for Astronomy, University of Amsterdam, Science Park 904, 1098 XH, Amsterdam, The Netherlands\\
$^{7}$Department of Astrophysics/IMAPP, Radboud University Nijmegen,
   P.O. Box 9010, 6500 GL Nijmegen, The Netherlands\\
$^{8}$Frankfurt Institute for Advanced Studies, Ruth-Moufang-Strasse 1, 60438 Frankfurt, Germany\\
$^{9}$School of Mathematics, Trinity College, Dublin 2, Ireland\\
}

\date{Accepted XXX. Received YYY; in original form ZZZ}

\pubyear{2020}
\usepackage{hyperref}
\hypersetup{
  linkcolor=blue,         
  citecolor=cyan,      
}

\begin{document}
\label{firstpage}
\pagerange{\pageref{firstpage}--\pageref{lastpage}}
\maketitle

\begin{abstract}
The Event Horizon Telescope (EHT) collaboration, an Earth-size
sub-millimetre radio interferometer, recently captured the first images
of the central supermassive black hole in M87. These images were
interpreted as gravitationally-lensed synchrotron emission from hot
plasma orbiting around the black hole.  In the accretion flows around
low-luminosity active galactic nuclei such as M87, electrons and ions are
not in thermal equilibrium. Therefore, the electron temperature, which is
important for the thermal synchrotron radiation at EHT frequencies of
230~GHz, is not independently determined. 
In this work, we investigate the commonly used parameterised
ion-to-electron temperature ratio prescription, the so-called R-$\beta$
model, considering images at 230~GHz by comparing with electron-heating
prescriptions obtained from general-relativistic magnetohydrodynamical
(GRMHD) simulations of magnetised accretion flows in a Magnetically
Arrested Disc (MAD) regime with different recipes for the electron
thermodynamics.  When comparing images at 230~GHz, we find a very good
match between images produced with the R-$\beta$ prescription and those
produced with the turbulent- and magnetic reconnection- heating
prescriptions. Indeed, this match is on average even better than that
obtained when comparing the set of images built with the R-$\beta$
prescription with either a randomly chosen image or with a time-averaged
one.
From this comparative study of different physical aspects, which include
the image, visibilities, broadband spectra, and light curves, we conclude
that, within the context of images at 230~GHz relative to MAD
accretion flows around supermassive black holes, the commonly-used and
simple R-$\beta$ model is able to reproduce well the various and more
complex electron-heating prescriptions considered here. 
\end{abstract}




\section{Introduction}
\label{sec:introduction}

High-frequency very-long-baseline interferometry (VLBI) on Earth-sized
baselines can resolve the immediate vicinity of nearby supermassive black
hole (SMBH) event horizons. The Event Horizon Telescope (EHT)
collaboration was established to build a global 1.3 mm-wavelength VLBI
network with the aim of capturing images of its primary targets:
Sagittarius A* (Sgr A*), the SMBH at the centre of our Galaxy, and
Messier 87 (M87), the active galactic nucleus (AGN) at the heart of the
Virgo A galaxy \citep{Doeleman2008,Goddi2017}. In April 2017, the EHT
made the first observations with a full array capable of imaging with all
eight participating radio telescopes, revealing an asymmetric ring
morphology of the central compact radio source in M87
\citep{Akiyama2019_L1, Akiyama2019_L2, Akiyama2019_L3, Akiyama2019_L4,
  Akiyama2019_L5, Akiyama2019_L6}. This image is interpreted as
gravitationally-lensed emission surrounding the black hole shadow
\citep{Akiyama2019_L1, Akiyama2019_L5}.

The mass-accretion rates of M87 and Sgr~A* are several orders of
magnitude less than the Eddington limit and, hence, the corresponding
luminosities of M87 and Sgr~A* are significantly lower than their
respective Eddington luminosities \citet[\eg][]{Ho2009,Prieto2016}.
Furthermore, recent Faraday-rotation measurements of Sgr~A* and M87 have
provided indirect evidence of low mass-accretion rates
\citet[\eg][]{Bower2003,Marrone2007,Kuo2014}. In this regime, material
accreting onto the central black hole is understood to be in the
radiatively inefficient accretion flow (RIAF) state, which comprises a
geometrically thick and optically thin accretion disc
\citep[\eg][]{Narayan1994,Yuan2014}. RIAF models have been employed to
investigate the innermost accretion flow structures for EHT target
objects via semi-analytic approaches
\cite[\eg][]{Broderick2006a,Broderick2009,Broderick2011,Broderick2016,Pu2016,Pu2018}.
In the past, many general-relativistic magnetohydrodynamical (GRMHD)
simulations have been performed for single-fluid RIAFs onto rotating
black holes for the study of event horizon-scale emission
\citep[\eg][]{Noble2007,Moscibrodzka2009,Moscibrodzka2012,Moscibrodzka2014,Moscibrodzka2016,Dexter2009b,Dexter2010,Shcherbakov2012,Chan2015b,Gold2017,Porth2017,Mizuno2018,Davelaar2018,Davelaar2019}.

In hot and low-density accretion flows such as RIAFs, Coulomb coupling
between electrons and ions is inefficient
\citep[\eg][]{Mahadevan1997,Mahadevan1998,Yuan2014}, and electrons and
ions are not in thermal equilibrium. In most single-fluid MHD
simulations, the ion temperature is dominant and therefore the electron
temperature, which is important for the radiation, cannot be determined
directly.

For modelling the emission from single-fluid GRMHD simulations, the
ion-to-electron temperature ratio is typically set manually in radiation
post-processing calculations. The simplest prescriptions for the electron
temperature take $T_{\rm i}/T_{\rm e}$ to be constant
\citep{Moscibrodzka2009}, dividing the simulation regions into jet and
disc components, with different temperature ratios in each region
\citep[\eg][]{Moscibrodzka2014,Chan2015b}. \cite{Moscibrodzka2016}
introduced a simple formula, the so-called ``R-$\beta$" prescription,
which is associated with plasma magnetisation. This ion-to-electron
temperature ratio prescription has been used in the development of
theoretical model observations in the image library of M87 by the EHT
\citep{Akiyama2019_L5}.  \cite{Anantua2020} have proposed several new
parameterised prescriptions as a function of the plasma beta or the
magnetic pressure, which are termed: critical beta electron temperature
model, constant electron beta model, and magnetic bias model. Typically,
a high ion-to-electron temperature ratio implies that the electron
heating does not impact the dynamics of the plasma flows because the
electron pressure is low.

Recently, \cite{Ressler2015} presented a new formulation of GRMHD
simulations which allows for the self-consistent evolution of the electron
fluid, including, independent of one another, the effects of electron heating
and conduction of heat flux along magnetic field lines.
This approach is applied to the modelling of Sgr A* images and spectra
\citep{Ressler2017}. \cite{Dexter2020} performed a parameter survey of
Sgr A* using different black-hole spins, electron-heating prescriptions,
and different accretion flow properties. \cite{Sadowski2017} and
\cite{Ryan2017} have extended this formulation to include the effects of
radiative feedback. This has been applied to investigate images and
variabilities of Sgr A* \citep{Chael2018} and M87
\citep{Ryan2018,Chael2019}.

From previous studies, we have two main approaches for the modelling of
ion-to-electron temperature ratio from GRMHD simulations: either setting
the ratio manually in the post-processing calculation, or calculating the
ratio from a more self-consistent evolution of the electron fluid from
GRMHD simulations. However, so far, direct comparison between these two
approaches has not been explored in detail. Consequently, in this
work we seek to compare the simplified R-$\beta$ model using 230~GHz
EHT images, time variability at 230~GHz, and the corresponding
broadband spectra with the results obtained from electron-heating
prescriptions of GRMHD simulations of accretion flows onto a black hole
with electron thermodynamics. In Sec.~\ref{sec:setup}, we present our
numerical approach and initial setup of GRMHD simulations and
general-relativistic radiative transfer (GRRT) calculations.

In this study, we focus on one accretion scenario, the Magnetically
Arrested Disc (MAD) \citep[\eg][]{Narayan2003,Tchekhovskoy2011}.  We show
our comparison results in 230\,GHz images, time variability of the
230\,GHz flux, and spectra from different black-hole spins, different
electron-heating prescriptions, and different inclination angles in
Sec.~\ref{sec:results}. In Sec.~\ref{sec:discussion}, we discuss our
findings and the limitations of our approach. We conclude in
Sec.~\ref{sec:conclusion}.

Throughout this paper, we adopt units where the speed of light, $c = 1$,
and the gravitational constant, $G = 1$. Self-gravity arising from the
gas is neglected. We absorb a factor of $\sqrt{4 \pi}$ into the
definition of the magnetic field 4-vector, $b^{\mu}$.

\section{Numerical Setup}
\label{sec:setup}

We have performed a set of three-dimensional (3D) GRMHD simulations of
magnetised tori in a black hole using the \texttt{BHAC} code
\citep{Porth2017, Olivares2019}. Simulations are initialised with a
Fishbone-Moncrief hydrodynamic equilibrium torus \citep{Fishbone76} with
parameters $r_{\rm in}= 20~r_{\rm g}$ and $r_{\rm max}= 40~r_{\rm g}$,
where $r_{\rm g} \equiv GM/c^2$ is the gravitational radius of the black
hole and $M$ is its mass.  An ideal-gas equation of state with a 
  constant relativistic adiabatic index of $\Gamma_{\rm g} = 4/3$ is
used \citep{Rezzolla_book:2013}. We note that some previous studies
  \citep{Sadowski2017,Chael2018,Chael2019} have used a variable equation
  of state in which the adiabatic index depends on the temperature.
This equilibrium torus solution is overlaid with a weak single magnetic
field loop, whose radial distribution of the field profile is designed to
supply enough magnetic flux onto the black hole to reach the magnetically
arrested disc (MAD) state \citep[\eg][]{Narayan2003,Tchekhovskoy2011}. In
order to excite the magneto-rotational instability (MRI) inside the
torus, 1\% of a random perturbation is applied to the gas pressure within
the torus. In this paper, we choose three values for the dimensionless
spin parameter: $a = -0.9375$, $0$, and $0.9375$.

The simulations are performed in spherical Modified Kerr-Schild
coordinates. The outer radial boundary is located at $r=2500\,M$.The
inner radial position of the simulation domain is well inside the black
hole horizon in all cases. The grid resolution is $384 \times 192 \times
192$, including the full $2 \pi$ azimuthal domain. The simulations are
evolved up to $t=15000\,M$ in order to reach a quasi-steady state.

We solve the electron thermodynamics during the evolution of single-MHD
fluid separately. For electron variables, we assume both charge
neutrality and that the electron number density and four-velocity are the
same as those of the ions, \ie $n_{\rm e} = n_{\rm i} = n$ and
$u^\mu_{\rm e} = u^\mu_{\rm i} = u^\mu$. However, the electron entropy
equation is solved for the electron temperature separately:
\begin{equation}
\rho T_{\rm e} \partial_\mu s_{\rm e} = f_{\rm e} Q \,,
\label{eq1}
\end{equation}
 where $\rho$ is the fluid rest-mass density, $s_{\rm
     e}=(\Gamma_{\rm e}-1)^{-1} \log(p_{\rm e}/\rho^{\Gamma_{\rm e}})$ is
   electron entropy, $\Gamma_{\rm e}$ is adiabatic index for the
   electrons, $p_{\rm e}$ is the electron pressure, $f_{\rm e}$ is a
 fraction of the dissipative heating which goes into electrons and $Q$ is
 the total heating rate per unit volume. Here we neglect the energy
 exchange rate due to Coulomb coupling, anisotropic thermal heat flux,
 and radiative cooling which have been considered in previous works
 \citep{Ressler2015, Ressler2017, Chael2018,
   Chael2019}.\footnote{\cite{Ressler2015} have included anisotropic
 conduction of heat along magnetic field lines. They reported that this
 conduction has little effect on the spectrum and image. \cite{Chael2018}
 have considered radiative cooling of electrons and Coulomb coupling of
 electrons to ions. These effects are mostly unimportant for low
 accretion rate systems such as Sgr A* and M87.}  The total heating rate,
 $Q$, is calculated through direct comparison between the internal energy
 obtained from solving an electron entropy conservation equation and the
 total internal energy of gas as described in \cite{Ressler2015}.
  
In order to apply the electron entropy equation in GRMHD simulations
using the \texttt{BHAC} code, eq.~(\ref{eq1}) has been rewritten as
\begin{equation}
\partial_\mu (\sqrt{-g} \rho u^\mu \kappa_{\rm e}) = \frac{\sqrt{-g}(\Gamma_{\rm e}-1)}{\rho^{\Gamma_{\rm e}-1}} f_{\rm e} Q \,,
\label{eq2}
\end{equation}
where $\kappa_{\rm e} \equiv \exp[(\Gamma_{\rm e} -1) s_{\rm e}]$.
Without the heat conduction term, equation (\ref{eq2}) takes a
conservative form, with conserved quantities $U_{\kappa_{\rm e}} \equiv
\sqrt{-g} \rho u^t \kappa_{\rm e}$ and flux $F^i_{\kappa_{\rm e}} \equiv
\sqrt{-g} \rho u^i \kappa_{\rm e}$. In order to obtain the time evolution
of $\kappa_{\rm e}$, we use operator splitting by following
\cite{Ressler2015}: (i) solve the conservative equation without a source
term, $S_{\kappa_{\rm e}}$, (ii) update $\kappa_{\rm e}$ with the heating
in the source term explicitly.

For the evaluation of heating in each time step, the entropy conversation
equation ($U_{\kappa_{\rm g}} \equiv \sqrt{-g} \rho u^t \kappa_{\rm g}$,
$F^i_{\kappa_{\rm g}} \equiv \sqrt{-g} \rho u^i \kappa_{\rm g}$, and
$S_{\kappa_{\rm g}}=0$) is introduced as a reference to compare with the
energy conservation equation. The heating update to the electrons is
calculated by the difference between the entropy obtained from the total
energy conservation equation ($\kappa_{\rm g}$) and the entropy obtained
from the entropy conservation equation ($\hat{\kappa}_{\rm g}$) as given
by
\begin{equation}
\kappa_{\rm e}^{n+1} = \hat{\kappa}^{n+1}_{\rm e} + \frac{\Gamma_{\rm e} -1}{\Gamma_{\rm g} -1} \left( \rho^{\Gamma_{\rm g} - \Gamma_{\rm e}} f_{\rm e} \right)^{n+1/2} (\kappa_{\rm g} - \hat{\kappa_{\rm g}})^{n+1} \,,
\label{eq3}
\end{equation}
where $\hat{\kappa}_{\rm e}$ is the solution from the electron entropy
conservation equation without a source term.
  
In GRMHD simulations, the heating is provided by grid-scale dissipation
which is related to magnetic reconnection, shock heating, Ohmic heating,
and turbulent heating. In this work, we consider two heating
prescriptions: turbulent heating and magnetic reconnection, to determine
the heating fraction $f_e$. In the turbulent heating model, we use the
results of numerical simulations of damping of MHD turbulence by
\cite{Kawazura2019}:
\begin{equation}
f_{\rm e} = \frac{1}{1+ Q_{\rm i}/Q_{\rm e}} \,,
\end{equation}
where
\begin{equation}
\frac{Q_{\rm i}}{Q_{\rm e}} = \frac{35}{1+(\beta/15)^{-1.4}~\exp \left(-0.1 T_{\rm e}/T_{\rm i}\right)} \,,
\end{equation}
and where $\beta \equiv p_{\rm g} / p_{\rm m}$ is the plasma-beta
parameter, the ratio between the fluid pressure and magnetic pressure
$p_{\rm m} = b^2/2$. This formula expresses a transition with
  increasing $\beta$ from an electron-dominated heating to a
  proton-dominated one; this transition takes place around the value of
  $\beta \sim 1$. \cite{Howes2010} has provided fitting formulae derived
from the linear theory of damping of MHD turbulence, showing
quantitatively similar behaviour to that described by
\cite{Kawazura2019}. Recently, \cite{Kawazura2020} extended their
investigation and provided new fitting formulae of the ion-to-electron
heating rate by turbulence, including the compressive-to-Alfv\'{e}nic driving
power ratio. In this work, we have not adopted this formula for turbulent
heating model.  In the magnetic reconnection model, we employ a fitting
function as measured in particle-in-cell simulations of magnetic
reconnection described by \cite{Rowan2017}:
\begin{equation}
f_{\rm e} = \frac{1}{2} \exp \left[ \frac{-(1-\beta/\beta_{\rm max})}{0.8+\sigma_{\rm h}^{0.5}} \right] \,,
\end{equation}
where $\beta_{\rm max} = \sigma_{\rm h}/4$, $\sigma_{\rm h} = b^2/\rho h$
is magnetisation as defined with respect to the fluid specific enthalpy
$h = 1 + \Gamma_{\rm g}~p_{\rm g} / (\Gamma_{\rm g}-1)$. In the
  formula for the magnetic-reconnection heating prescription, $f_{\rm e}$
  reaches its maximum value, \ie 1/2, in highly magnetised regions
  ($\sigma_{\rm h} \gg 1$) or in regions with large $\beta$. On the other
  hand, in regions with small $\beta$ ($\beta \ll \beta_{\rm max}$),
  $f_{\rm e}$ attains small values that depend on $\sigma_{\rm
    h}$. Finally, in the limit of non-relativistic reconnection, \ie
  $\sigma_{\rm h} \ll 0.1$, $f_{\rm e} \rightarrow 0.14$.  Recently
\cite{Rowan2019} has provided a new fitting formula for the heating
prescription for the magnetic reconnection model, including the
effect of guide fields. A quantitatively similar behaviour to that found
in \cite{Rowan2017} was demonstrated. We here perform GRMHD
  simulations considering two different electron heating prescription --
  either turbulent or magnetic-reconnection heating -- and three
  different black-hole spins, for a total six 3D GRMHD simulations.

In the GRMHD simulations, we take the adiabatic index of the electron
fluid to be $\Gamma_{\rm e} = 4/3$ and initially set the electron
internal energy density $u_{\rm e}$ to 10\% of the fluid internal energy
density $u_{\rm g}$. The evaluation of the implementation of the
electron-heating calculations in the \texttt{BHAC} code and its
convergence is presented in Appendix~\ref{sec:shock_test}, where we
  show its correct implementation and convergence when $\Gamma_{\rm g} =
  \Gamma_{\rm e}$.

As customary in codes solving the equations of relativistic
  hydrodynamics or those of GRMHD, the occurrence of vacuum is avoided by
  introducing a very low-density fluid, \ie an atmosphere, filling
  regions that are far from the high-density fluid
  \citep{Rezzolla_book:2013}. In essence, floor values are applied to the
  rest-mass density, $\rho_{\rm fl} = 10^{-4} r^{-2}$ and the gas
  pressure, $p_{\rm fl}=(10^{-6}/3) r^{-2\Gamma_{\rm g}}$, so that in all
  the numerical cells for which $\rho \le \rho_{\rm fl}$ or $p \le p_{\rm
    fl}$, we simply set $\rho = \rho_{\rm fl}$ and $p=p_{\rm
    fl}$. Similarly, a ceiling is introduced in those regions of high
  magnetisation, so that we set $\sigma_{\rm max}=100$ in all those cells
  for which $\sigma \geq \sigma_{\rm max}$. Finally, for the electron
  entropy we apply both a floor and a ceiling to the electron
  pressure. More specifically, if the electron pressure is less than
  $1\%$ of the floor value of the gas pressure $p_{\rm fl}$, we then
  reset $p_{\rm e} = 0.01 p_{\rm fl}$. At the same time, if the electron
  pressure is larger than the gas pressure, we reset $p_{\rm e} = 0.99
  p_{\rm g}$. We note that we re-calculate the gas entropy $\kappa_{\rm
    g}$ and the electron entropy $\kappa_{\rm e}$ in those cells where
  the density is floored.

In order to obtain the synthetic image from GRMHD simulations, we perform
GRRT calculations in post-processing using \texttt{BHOSS}
\citep{Younsi2012,Younsi2020a}. The \texttt{BHOSS} code performs GRRT
calculations of the GRMHD simulation data in post-processing.  The
equations of GRRT are solved along geodesics integrated through this
GRMHD data, and the resultant images, light curves and spectra as seen by
a distant observer for a given viewing angle and observing frequency are
determined.

In this work, we adopt a relativistic thermal Maxwell-J\"unttner electron
distribution function for the synchrotron absorption and emission given
by \cite{Leung2011}. For the electron-ion temperature ratio, we consider
two different approaches. The first uses the aforementioned R-$\beta$
prescription:
\begin{equation}
\frac{T_{\rm i}}{T_{\rm e}} = \frac{1}{1 + \beta^2} R_{\rm l} + \frac{\beta^2}{1+\beta^2} R_{\rm h} \,.
\end{equation}
In the R-$\beta$ prescription, the temperature ratio in the strongly
magnetised regions ($\beta_p \ll 1$) like the jet funnel is $T_{\rm
  i}/T_{\rm e} \sim R_{\rm l}$ and in weakly magnetised regions ($\beta_p
\ll 1$), such as the disc, this tends to $T_{\rm i}/T_{\rm e} \sim R_{\rm
  h}$.  In our work, we fix $R_{\rm l} = 1$ and vary $R_{\rm h}$ as
$R_{\rm h}= 1$, $5$, $10$, $20$, $40$, $80$, and $160$, a similar choice
to that of \cite{Akiyama2019_L5}.  Here, we keep $R_\mathrm{l}=1$
  fixed and vary $R_\mathrm{h}$ from 1 to 160, following the analysis
  carried out by \cite{Akiyama2019_L5}. Nevertheless, in
Appendix~\ref{sec:diff_Rlow}, we investigate the effect of adopting
different values for $R_\mathrm{l}$ and conclude that the
image-comparison results do not depend on the choice made for
$R_\mathrm{l}$.

The second approach takes the ion-to-electron temperature ratio directly
from the electron-heating prescription calculated in the GRMHD
simulations.  The electron pressure is computed as $p_{\rm e} =
  \kappa_{\rm e} \rho^{\Gamma_{\rm e}}$, where the values of $\kappa_{\rm
    e}$ are updated as in Eq. \eqref{eq3}. The ion-to-electron
  temperature ratio is then calculated as $T_{\rm i} / T_{\rm e} =
  (p_{\rm g} - p_{\rm e}) / p_{\rm e}$.
 
The dimensionless electron temperature is then given by:
\begin{equation}
  \Theta_{\rm e} = \left(\frac{p_{\rm g}-p_{\rm e}}{\rho} \right)
  \left( \frac{m_{\rm p}/m_{\rm e}}{T_{\rm i}/T_{\rm e}}\right) \,,
\end{equation}
where $m_{\rm p}$ and $m_{\rm e}$ are the proton and electron masses,
respectively, and we assume a fully ionised hydrogen
plasma.\footnote{ In the R-$\beta$ parameterisation, we use $p_{\rm
    g}$ instead of $p_{\rm g}-p_{\rm e}$ because it is not possible to
  estimate $p_{\rm e}$ from one-temperature GRMHD simulations. This is
  equivalent to assuming $T_{\rm i} \simeq T_{\rm g}$ and $T_{\rm i} \gg
  T_{\rm e}$ \citep[e.g.,][]{Moscibrodzka2016}.} The electron
temperature in c.g.s.~units is calculated as $T_{\rm e} = m_{\rm e} c^2
\Theta_{\rm e}/k_{\rm B}$, where $k_{\rm B}$ is the Boltzmann constant.

Although GRMHD simulations are scale-free, the GRRT calculation depends
on the physical mass scale. Here we consider our target central black
hole to be M87 with a mass $M= 6.5 \times 10^9 M_\odot$ at a distance $D
= 16.8\,\mathrm{Mpc}$ \citep{Akiyama2019_L6}. We set the mass scale unit
which is the conversion from simulation density to physical density by
normalising the time-averaged flux at 230\,GHz to the value of 0.5\,Jy
\citep{Akiyama2019_L5}.

In GRRT calculations, we limit the emission regions in which we can
reasonably trust the fluid thermodynamics by imposing a threshold on the
magnetisation $\sigma = b^2/\rho$. We only consider emission that
originates from the regions where $\sigma < 1$. In regions with $\sigma >
1$, the fluid density and temperature may be affected by the simulation
floors and are therefore unreliable. Furthermore, we also adopt the `fast
light' approximation. In this approximation, we calculate a emission on a
fixed time slice of fluid quantities which means the light propagation
time across the domain is small compared to the dynamical timescale of
the system.

\section{Results}
\label{sec:results}
\subsection{GRMHD Simulations with Electron Thermodynamics}
\label{GRMHD}
\begin{figure}
\begin{center}
 \includegraphics[width=\linewidth]{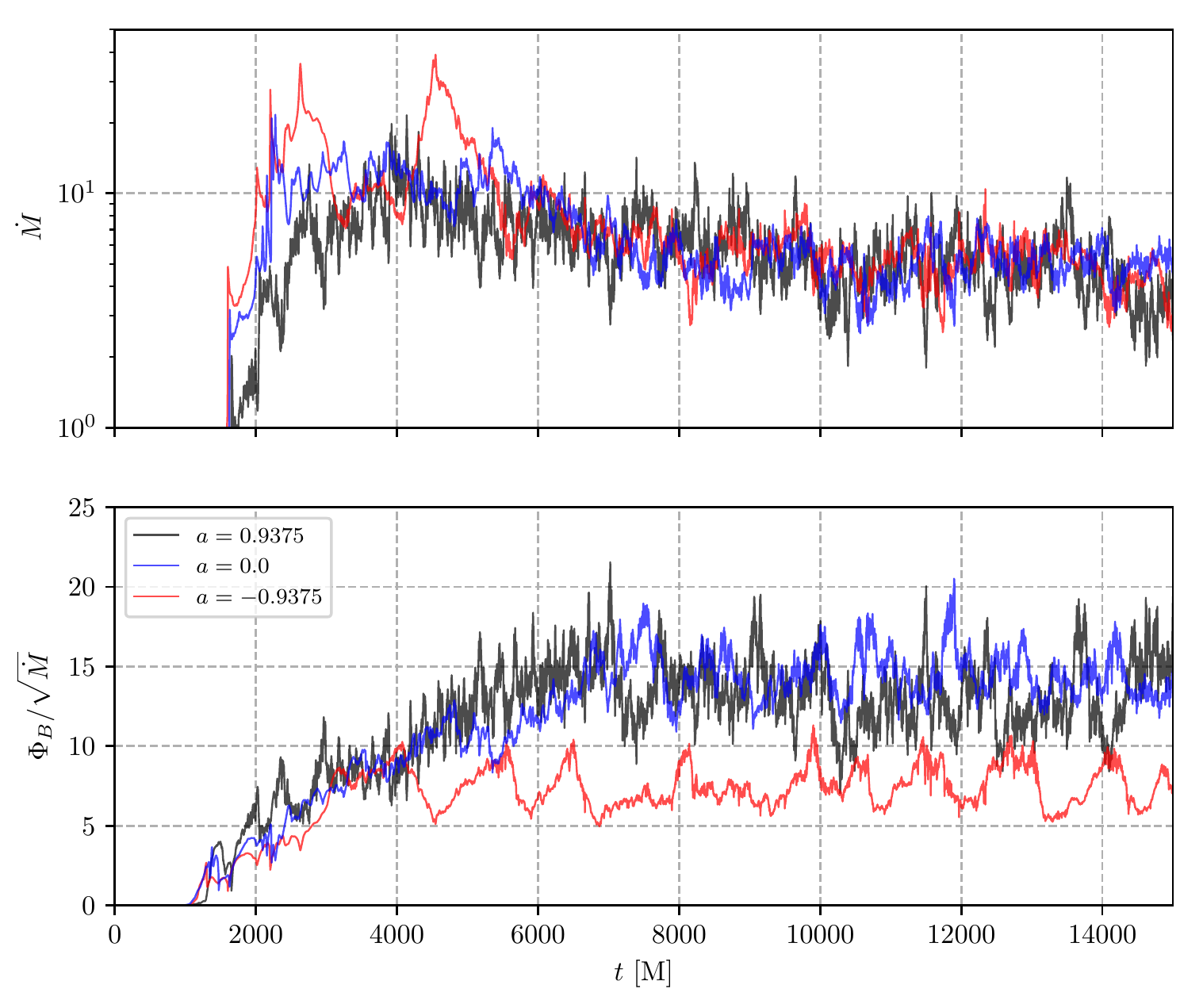}
 \caption{({\it Top}) mass-accretion rate and ({\it bottom}) magnetic
   flux rate at the black hole horizon for a black hole with $a=-0.9375$
   (red), $0$ (blue), and $0.9375$ (black).}
 \label{fig:Mdot}
\end{center}
\end{figure}

\begin{figure*}
\begin{center}
 \includegraphics[width=0.33\linewidth]{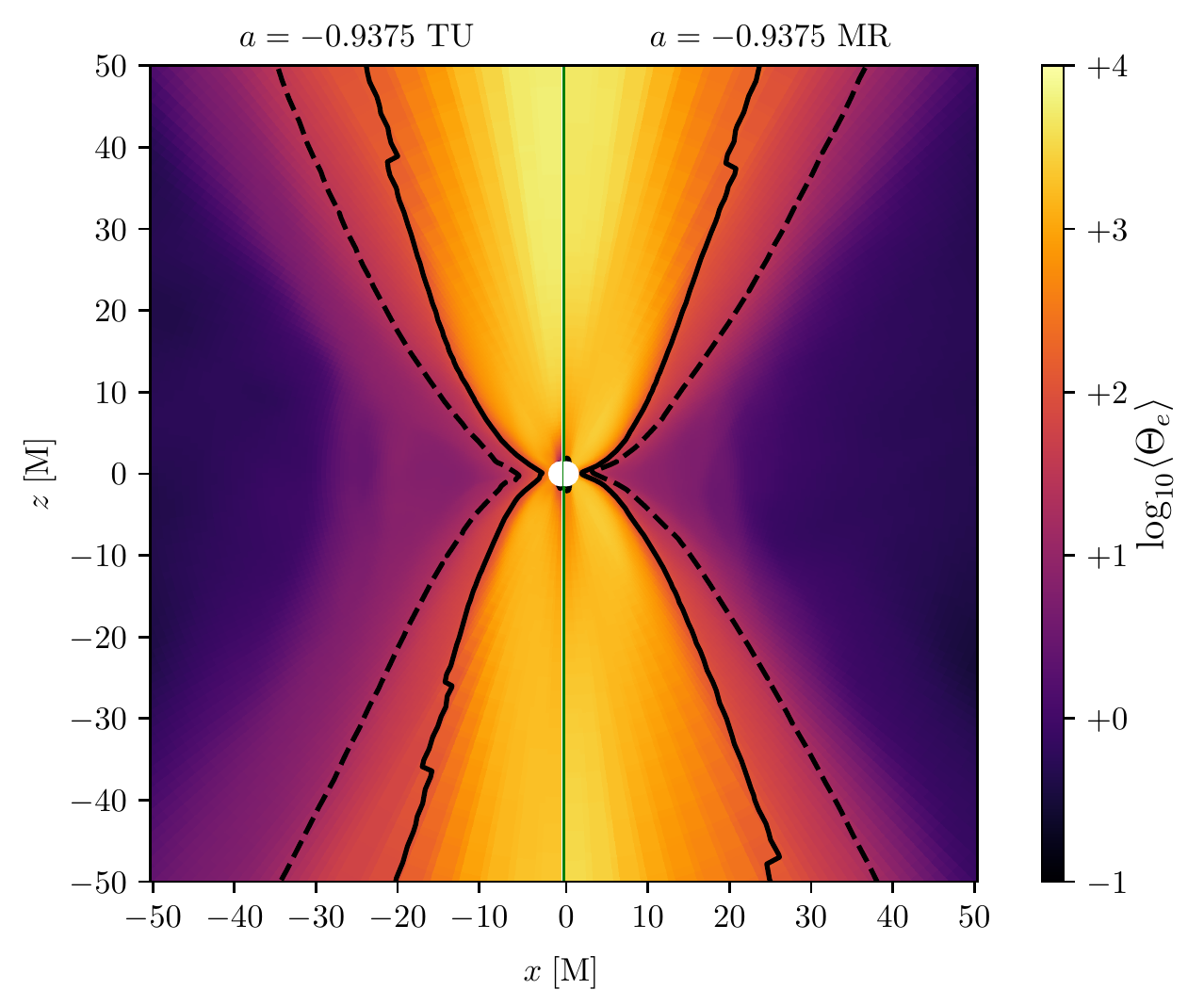}
 \includegraphics[width=0.33\linewidth]{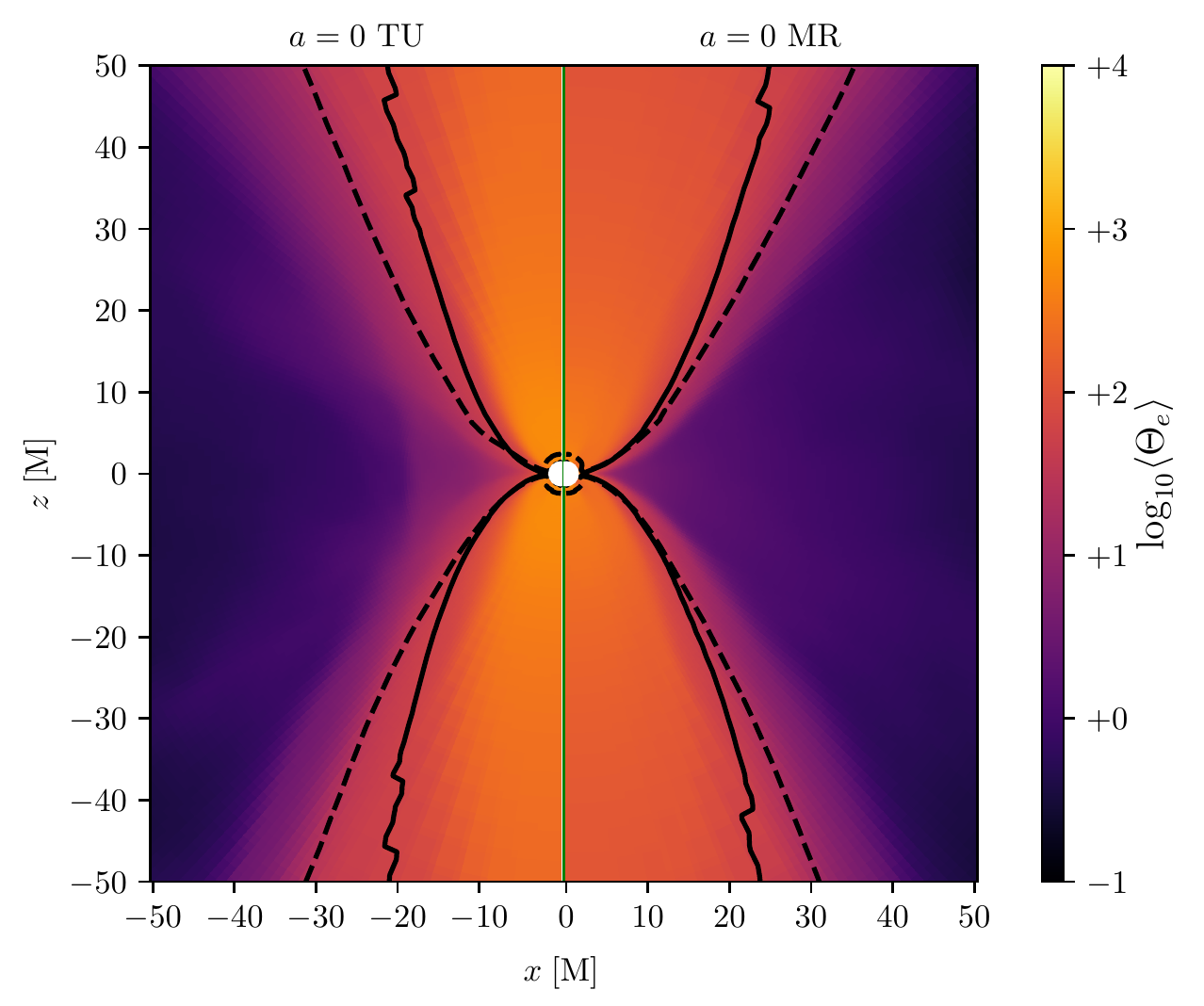}
 \includegraphics[width=0.33\linewidth]{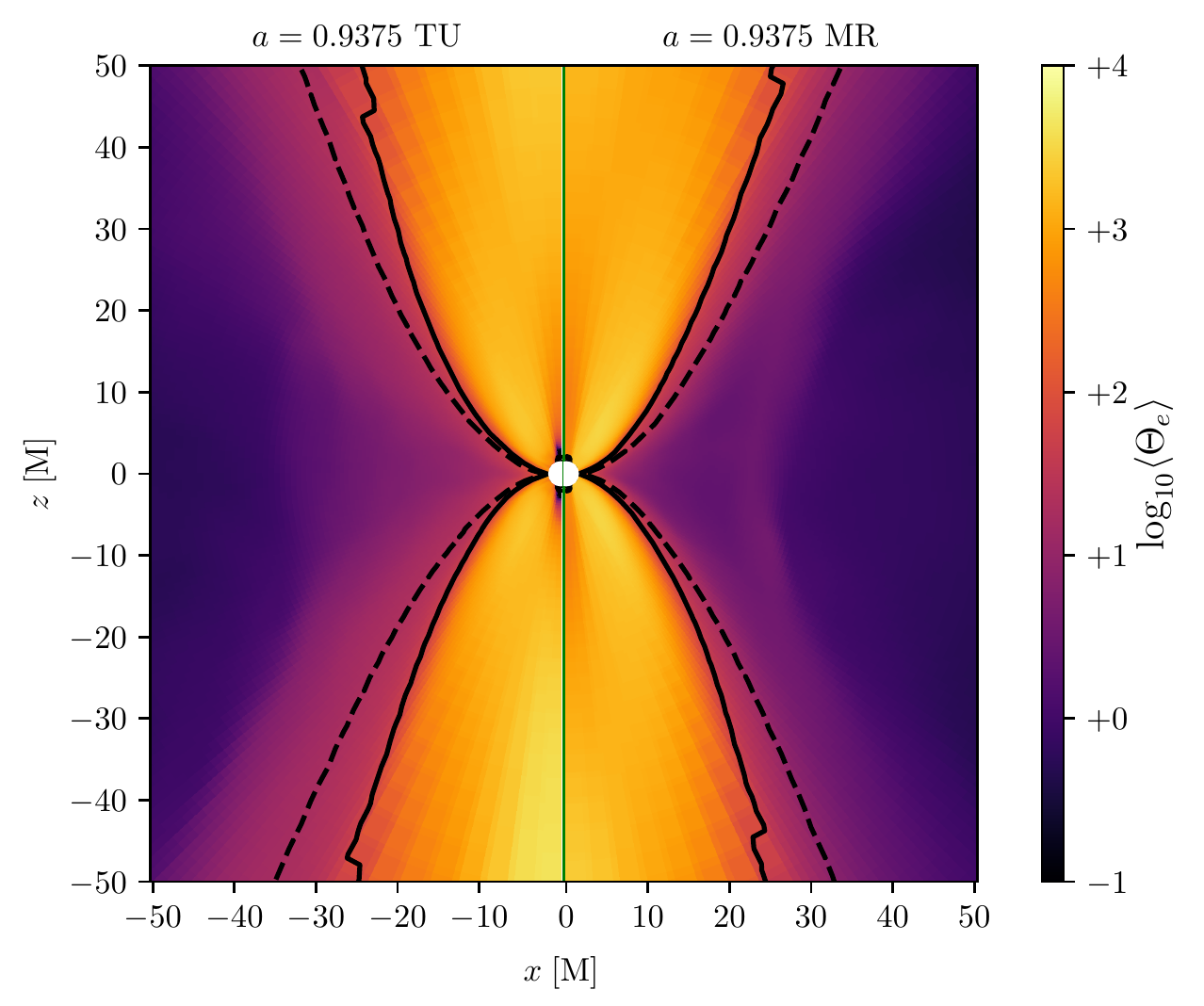}
 \caption{Azimuthal and time averaged dimensionless electron temperature
   $\Theta_e$ for black-hole spins with $a=-0.9375$ ({\it left}), $0$
   ({\it middle}), and $0.9375$ ({\it right}) using the turbulent
   heating prescription ({\it left side of the panels}) and the
   magnetic-reconnection heating prescription ({\it right side of the
     panels}). The averaging is performed over the time interval $t=14000
   - 15000 M$, which is when the simulations have reached a quasi-steady
   state. The black solid and dotted lines indicate $\sigma =1$ and
   Bernoulli parameter $-hu_t = 1.02$, respectively.}
 \label{fig:GRMHD_Thetae}
\end{center}
\end{figure*}

\begin{figure*}
\begin{center}
 \includegraphics[width=0.33\linewidth]{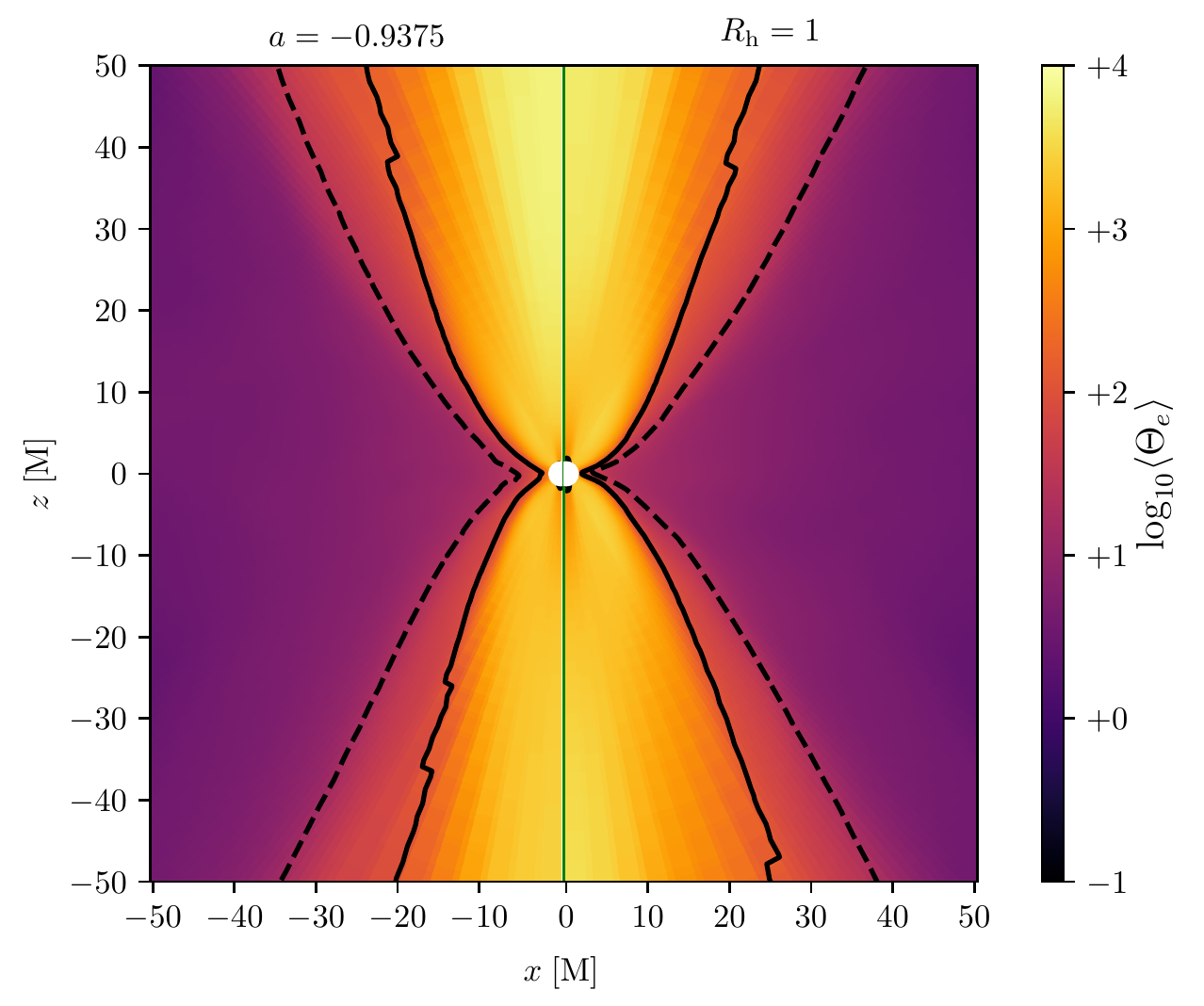}
 \includegraphics[width=0.33\linewidth]{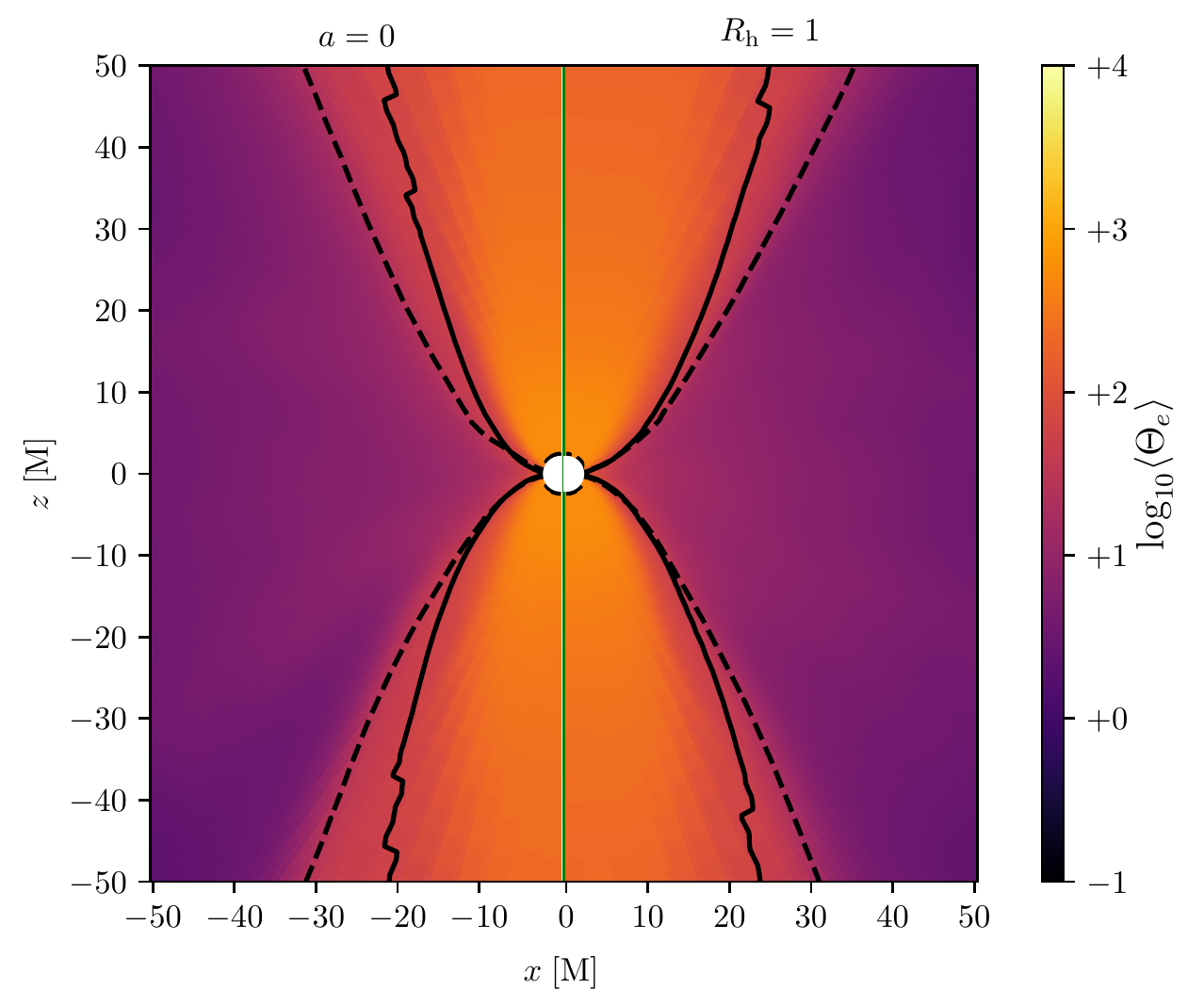}
 \includegraphics[width=0.33\linewidth]{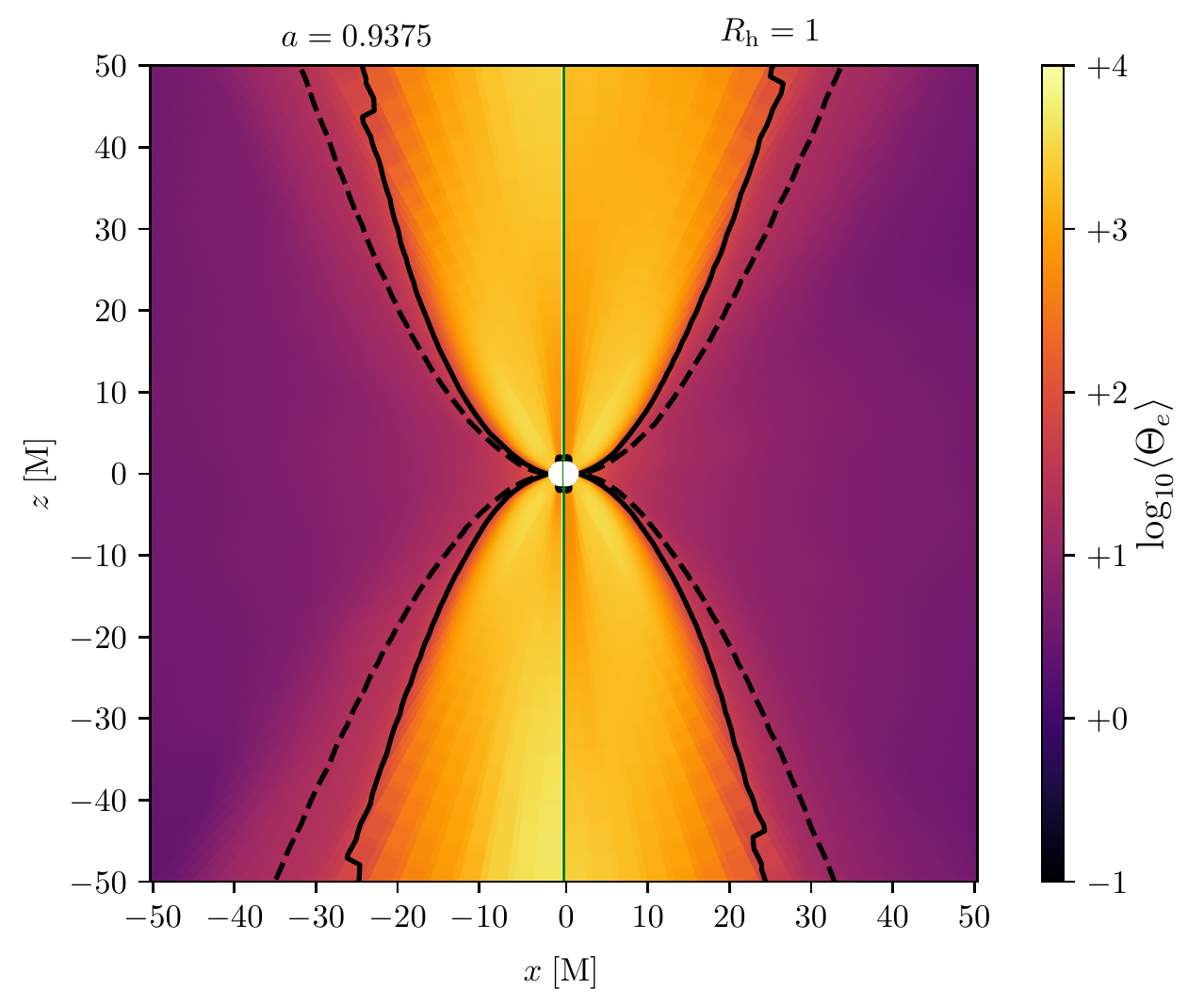}
 \includegraphics[width=0.33\linewidth]{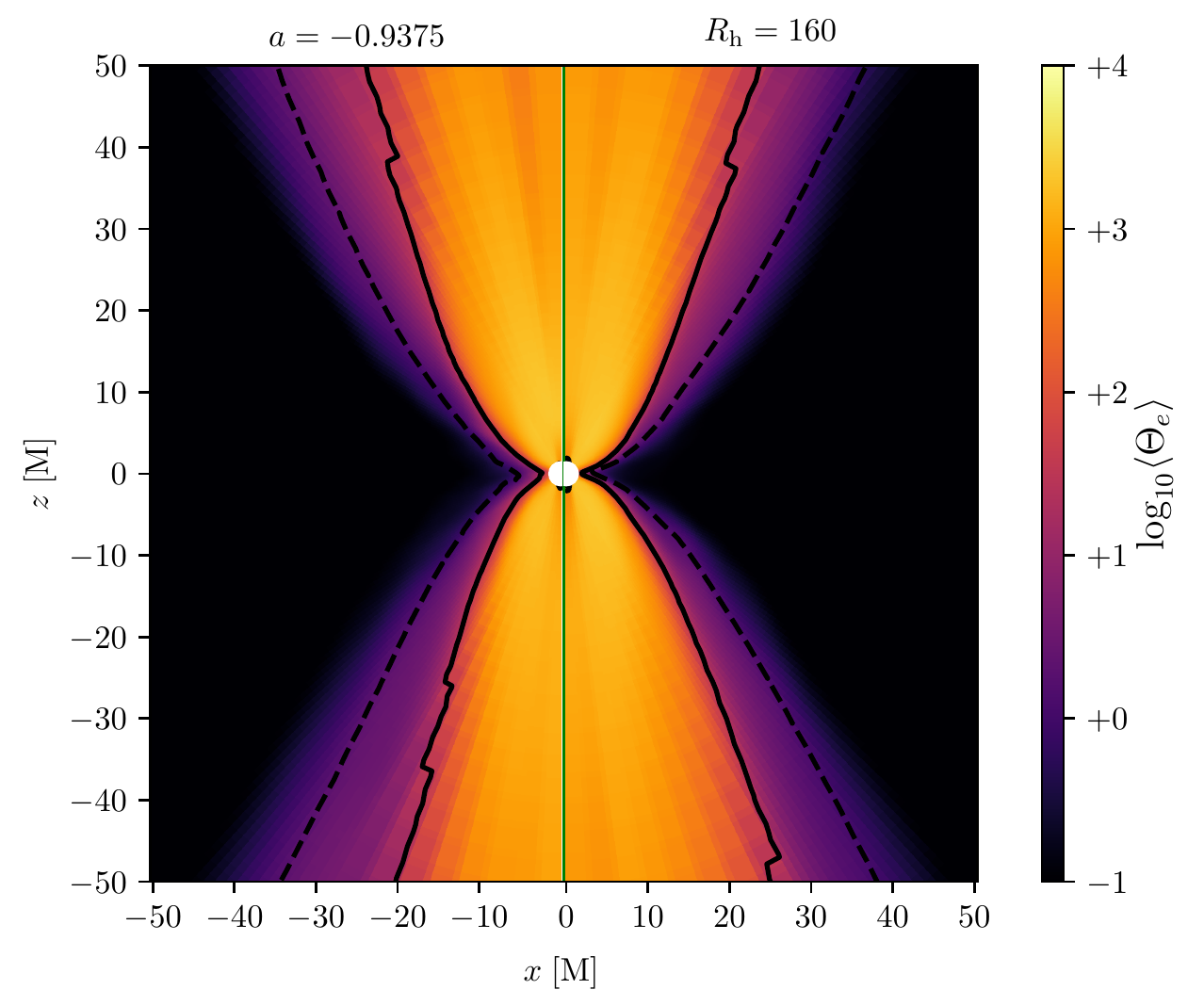}
 \includegraphics[width=0.33\linewidth]{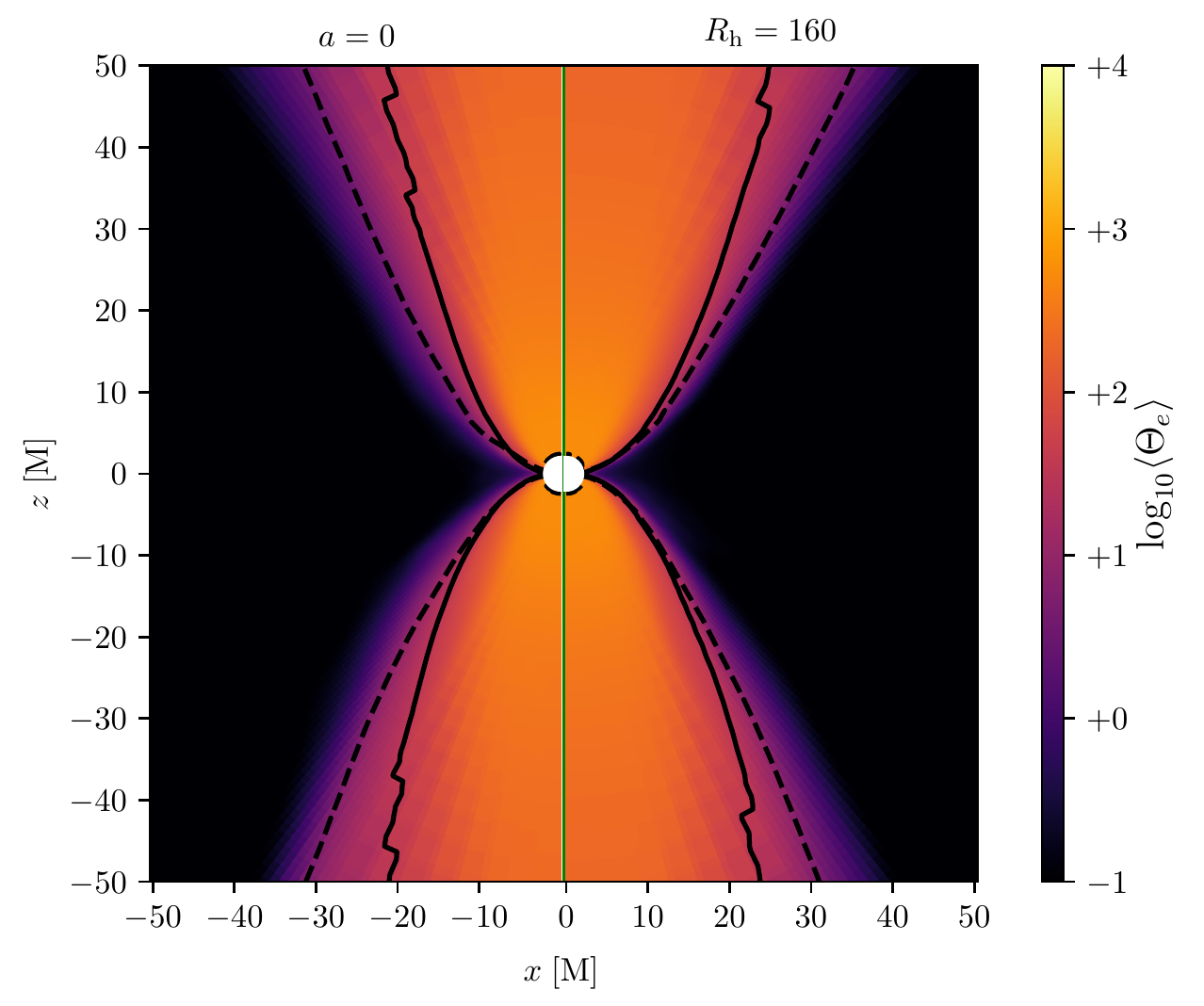}  
 \includegraphics[width=0.33\linewidth]{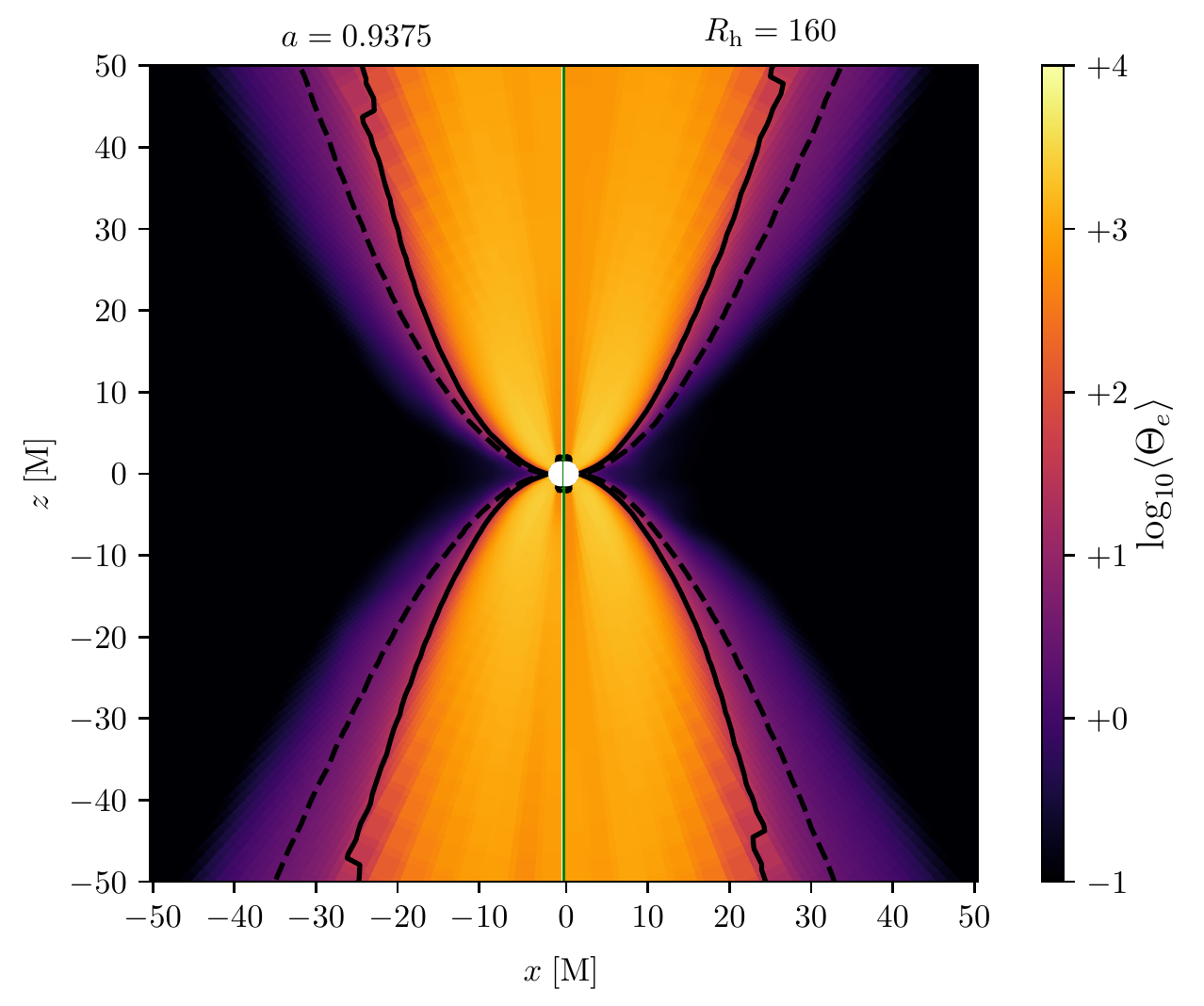}
 \caption{Same as Fig.\ref{fig:GRMHD_Thetae}, but when using
     $R_{\rm h}=1$ ({\it upper}) and $R_{\rm h}=160$ ({\it lower}) in the
     R-$\beta$ parameterised prescription.}
 \label{fig:GRMHD_ThetaeM}
\end{center}
\end{figure*}

In order to investigate the properties of accretion flows quantitatively,
we calculate the volume-integrated mass-accretion rate and magnetic flux
rate at the black hole horizon. Following \cite{Porth2019} we define the
mass-accretion rate as
\begin{equation}
\dot{M} = \int^{2\pi}_0 \int^\pi_0 \rho u^r \sqrt{-g} d \theta d\phi,
\end{equation}
while the magnetic flux rate is written as
\begin{equation}
\Phi_\mathrm{B} = \frac{1}{2} \int^{2\pi}_0 \int^\pi_0  |B^r| \sqrt{-g} d \theta d\phi.
\end{equation}
Figure \ref{fig:Mdot} shows time evolution of mass-accretion rate
($\dot{M}$) and dimensionless magnetic flux rate at the black hole
horizon ($\dot{\phi_\mathrm{B}}/\sqrt{\dot{M}} $) in black-hole spin
cases with $a=-0.9375$ (red), $0$ (blue), and $0.9375$ (black). The
mass-accretion rates have very similar profiles amongst different
black-hole spin cases, which gradually decrease with time after
$t=4000\,M$. The dimensionless magnetic flux saturates at the maximum
values of $\Phi_\mathrm{B}/\sqrt{\dot{M}} \simeq 15$ in the cases with
$a=0.9375$ and $0$. In counter-rotating case with $a=-0.9375$, the
maximum values of dimensionless magnetic flux adopts a somehow lower
value of 8. These values are consistent with \cite{Tchekhovskoy2012}.
\footnote{In our system of units, the dimensionless magnetic flux rate at
the black hole horizon differs from the used definition in
\cite{Tchekhovskoy2011,Tchekhovskoy2012,McKinney2012} by a factor of
$\sqrt{4\pi}$.  }  All cases reach the MAD state after $t=6000\,M$. 
  As first discussed by \citet{Tchekhovskoy2012}, a counter-rotating
  accretion flow leads generically to a smaller saturation value of the
  magnetic flux than in the case of a co-rotating flow. Furthermore, the
  results obtained here for the magnetic flux are systematically smaller
  than those reported by \citet{Dexter2020} for co-rotating flows,
  hinting that the actual saturation value may depend on the
  initial disc thickness, which is different from the one adopted by
  \citet{Dexter2020}.

Azimuthal- and time- averaged dimensionless electron temperatures
($\Theta_e$) are shown in Figure~\ref{fig:GRMHD_Thetae}.  In the funnel
wall region between the highly magnetised polar funnel and bound disc
material, electrons are efficiently heated in both heating prescriptions,
resulting in a high ion-to-electron temperature ratio. This trend is
similarly seen in \cite{Chael2019} and \cite{Dexter2020}, which also used
MAD simulations.  We note that our MAD simulations exhibit efficient
heating in the polar and funnel wall regions. This is due to the
difference in the assumed adiabatic index of the fluid between GRMHD
simulations. In our simulations, we assume an identical adiabatic index
between fluid and electrons, \ie $\Gamma_{\rm g}=\Gamma_{\rm e}=4/3$. In
this case, electron heating occurs more efficiently via dissipation (see
Appendix~\ref{sec:shock_test}). In our simulations, we neglect radiative
cooling. As a consequence, this may lead to an overestimate in cases of
high electron temperature.

\begin{figure}
\begin{center}
\includegraphics[width=0.9\linewidth]{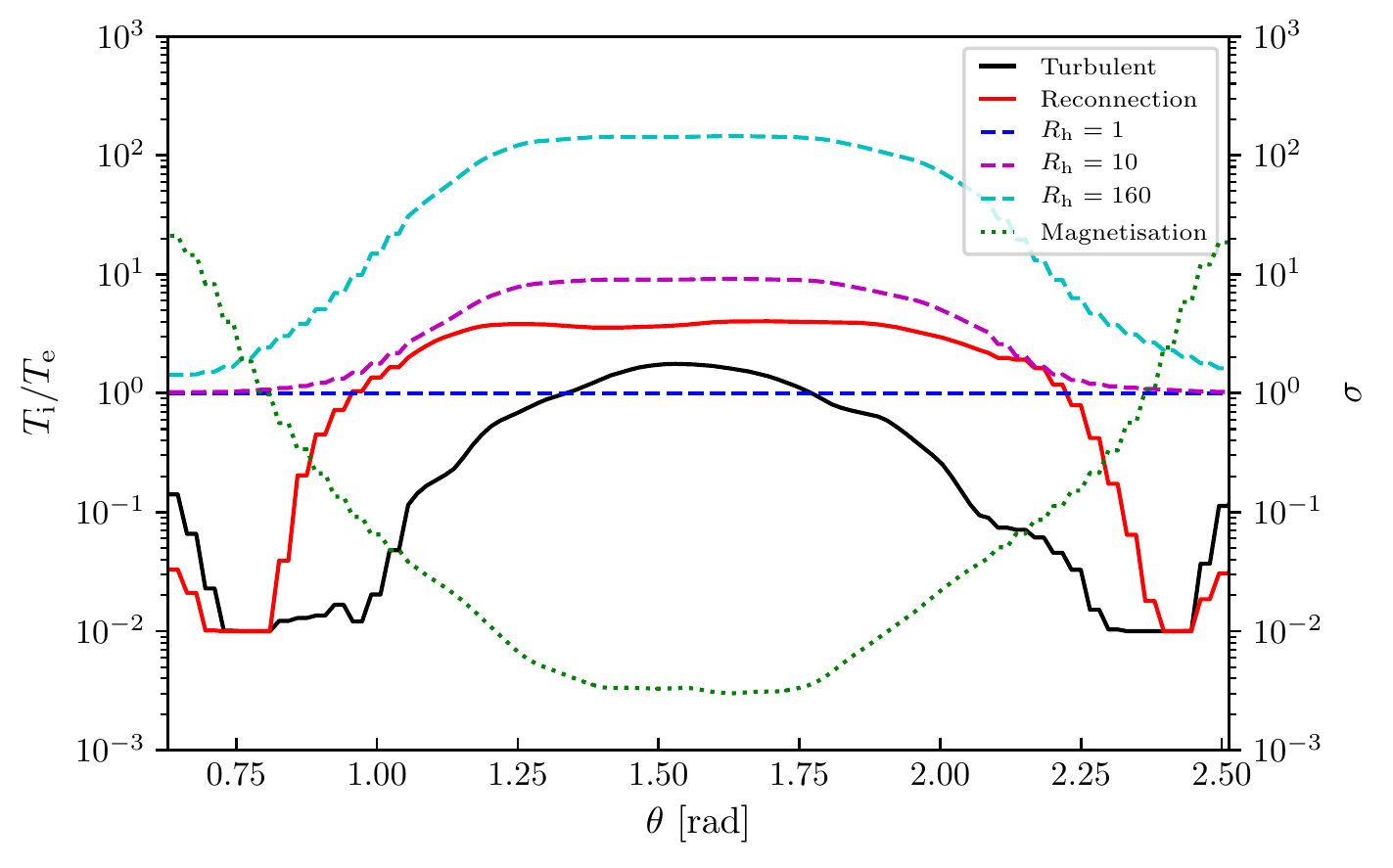}
\caption{Azimuthal and time-averaged polar-angle profiles of the
  ion-to-electron temperature ratio of the MAD-accretion case with
  $a=0.9375$ at $r=20\,M$. Solid lines indicate turbulent (black) and
  magnetic reconnection (red) heating prescription. Dashed lines present
  R-$\beta$ parameterised prescription with different $R_\mathrm{h}$
  value, $R_\mathrm{h}=1$ (blue), 10 (magenta), and 160 (cyan). The green
  dotted line shows the magnetisation.}
 \label{fig:GRMHD_theta_TiTe}
\end{center}
\end{figure}

As reported by \cite{Dexter2020}, we also see a mild dependence on
black-hole spin in the electron temperature, which becomes larger for
higher black-hole spins. However, both the ion and the electron
temperatures increase, so that the ion-to-electron temperature ratio does
not vary appreciably.  As a reference, we present the azimuthal- and
time- averaged dimensionless electron temperature by using the R-$\beta$
prescription with $R_{\rm h} =1$ and $R_{\rm h} = 160$ in
Fig.~\ref{fig:GRMHD_ThetaeM}. The dimensionless electron temperature in
the disc region becomes cooler for higher values of $R_{\rm h}$.

We note that the region near the poles is highly magnetised and may be
affected by the numerical floor treatment. Therefore the ion-to-electron
temperature ratio may be unreliable in this region and we omit the
contribution of these regions in the GRRT calculations of images at
230\,GHz (the threshold is set with $\sigma\ge1$).

 Figure~\ref{fig:GRMHD_theta_TiTe} presents azimuthal- and time-
  averaged polar-angle profiles of the ion-to-electron temperature ratio
  of the MAD model with $a=0.9375$ on a polar slice at $r=20M$. Both the
  R-$\beta$ (dashed lines) and the electron-heating prescriptions (solid
  lines) produce a high ion-to-electron temperature ratio around the
  equatorial plane, although the R-$\beta$ generally yield larger
  temperatures than the reconnection-heating or the turbulent-heating
  prescriptions; similar plots have been shown by \citet{Chael2018}. We
  note that the reconnection-heating prescription employed by
  \citet{Chael2018} leads to $T_{\rm e} < T_{\rm i}$ everywhere along the
  polar section, while in our calculations this is true only in the
  highly magnetised region.  Because in this region radiative cooling
  would be effective, the electron heating computed in our simulations
  may be overestimated. Also, it should be noted that around the
  equatorial plane, both electron-heating prescription have a profile
  similar to that of the R-$\beta$ prescriptions. Large differences are
  seen around the funnel region, where most of the electron heating takes
  place, so that $T_{\rm i} \ll T_{\rm e}$ there for both
  electron-heating prescriptions.

\subsection{GRRT Calculations}
\begin{figure*}
\begin{center}
 \includegraphics[width=0.7\linewidth]{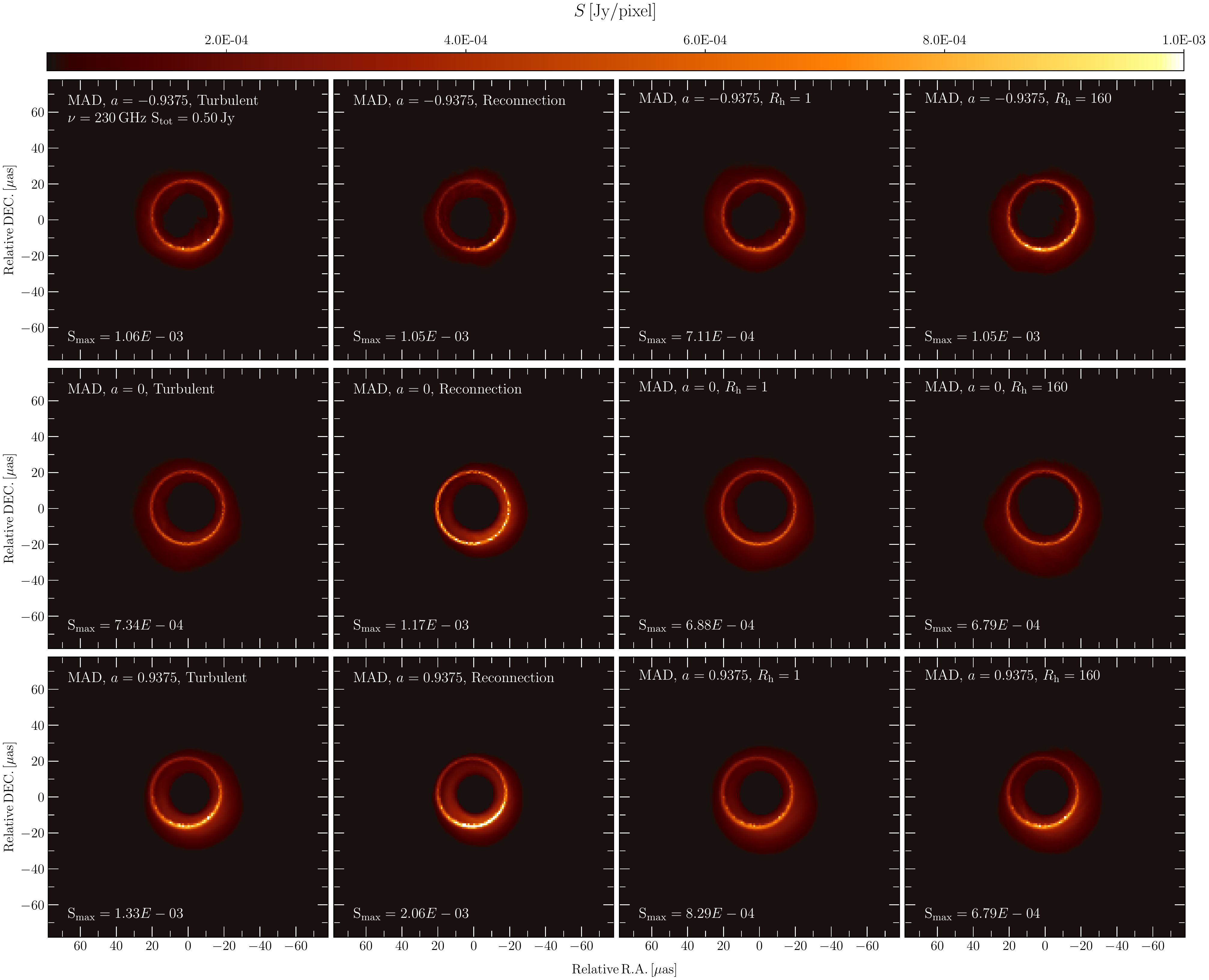}
 \caption{Time-averaged GRRT images at $i=163\degr$ of MAD simulations
   with different black-hole spin cases, $a=-0.9375$ (top panels), $0$
   (middle panels), and $0.9375$ (top panels). From left to right: images
   using turbulent heating prescription, magnetic-reconnection heating
   prescription, the R-$\beta$ model with $R_\mathrm{h}=1$, and R-$\beta$
   model with $R_\mathrm{h}=160$. The image is averaged with GRRT images
   from $t=14000\,M$ to $15000\,M$. All averaged images have the same total
   flux with 0.5\,Jy at 230\,GHz.}
 \label{fig:GRRT_i163_lin}
\end{center}
\end{figure*}

\begin{figure*}
\begin{center}
 \includegraphics[width=0.7\linewidth]{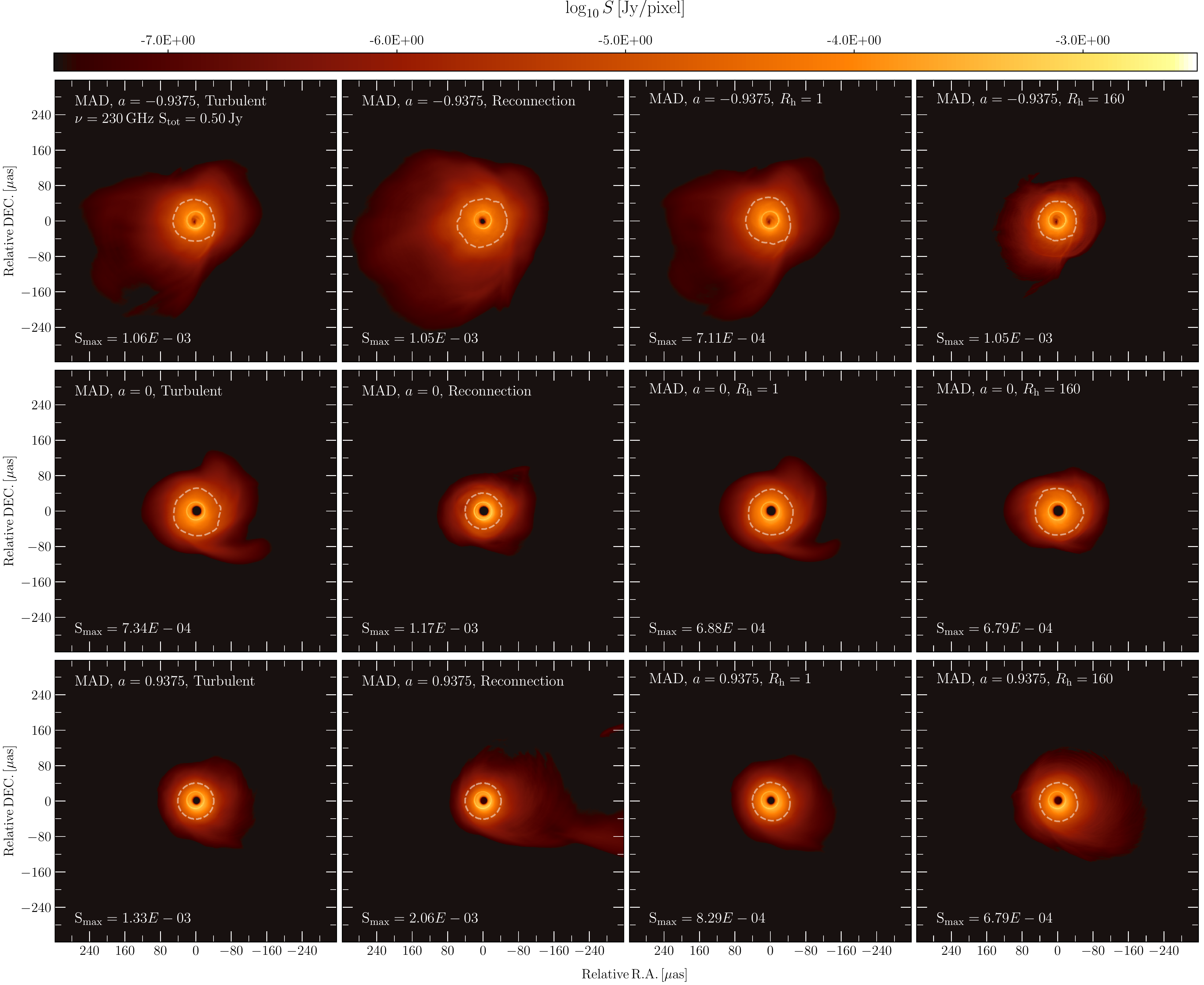}
 \caption{Same as Fig. \ref{fig:GRRT_i163_lin}, but plotting logarithmic
   scale. Dashed lines are indicated the 1\% of maximum intensity in each
   images.}
 \label{fig:GRRT_i163_log}
\end{center}
\end{figure*}

In order to compare radiative signatures using different ion-to-electron
temperature ratio prescriptions, we have calculated 230\,GHz images using
the GRRT code \texttt{BHOSS} \citep{Younsi2012,Younsi2020a}.
Figures~\ref{fig:GRRT_i163_lin} and \ref{fig:GRRT_i163_log} show the
time-averaged GRRT images at $i=163\degr$ of MAD simulations with
different black-hole spins: $a=-0.9375$ (top panels), $0$ (middle
panels), and $0.9375$ (bottom panels) using the turbulent heating
prescription, the magnetic-reconnection heating prescription, and the
R-$\beta$ model with $R_\mathrm{h}=1$ and $R_\mathrm{h}=160$, in a linear
and logarithmic scale, respectively. Images are averaged from $t=14000\,M$
to $15000\,M$. All averaged images have the same total flux of 0.5\,Jy. The
camera field of view is set to be 640 $\mu$as $\times$ 640 $\mu$as, which
is a four times larger field-of-view than the one used by the
\cite{Akiyama2019_L5}.  For the black hole in M87, 100 $r_\mathrm{g}$
corresponds to 382 $\mu$as. Our choice of field of view corresponds to
167 $r_\mathrm{g}$. We rotate the image to 252\degr (72\degr)
North-to-East in co-rotating case (counter-rotating case) to set the
bright spot position to be in the South. In the co-rotating case, the
approaching jet orientation is in a similar direction to the large scale
jet \citep[\eg][]{Hada2017,Kim2018,Walker2018}.

In the time-averaged images, we effectively suppress the time-dependent
turbulent features of the individual GRRT images, thus allowing for more
generic and persistent features of the GRRT images to emerge.  In high
black-hole spin cases, the asymmetry of the brightness distribution of
the bright photon ring becomes stronger than the non-rotating black hole
case, which is almost uniform.  GRRT images at 230\,GHz in different
ion-to-electron temperature ratio prescriptions are morphologically very
similar, \ie bright photon ring emission with some faint extended
emission around it. However, the brightness distribution of the photon
ring differs with the chosen heating prescriptions. In the turbulent
heating prescription, the brightness distribution of the photon ring is
more uniform than the magnetic reconnection heating prescription. A
similar trend is seen for higher $R_\mathrm{h}$ values in the R-$\beta$
model. In the images with a logarithmic scale, these trends are more
clear. Both turbulent and magnetic heating prescriptions have more
extended diffused emission than the R-$\beta$ models. This is because both
electron-heating prescriptions have higher ion-to-electron temperature
ratios at the funnel wall region. In the turbulent heating prescription
model, the maximum flux at 230\,GHz is lower than other models due to
contributions from large extended diffused emission regions. This leads to
a region representing the 1\% level of the maximum flux (white dashed
contour in Fig.~\ref{fig:GRRT_i163_log}) that is a little larger than in the
$R-\beta$ models. On the other hand, the magnetic-reconnection heating
prescription model has a higher maximum flux even though it has a more
diffused extended emission region, which may not contribute much to the
total flux. This is indicated by the position of the 1\% level of maximum
flux.
 
\begin{figure*}
\begin{center}
 \includegraphics[width=0.7\linewidth]{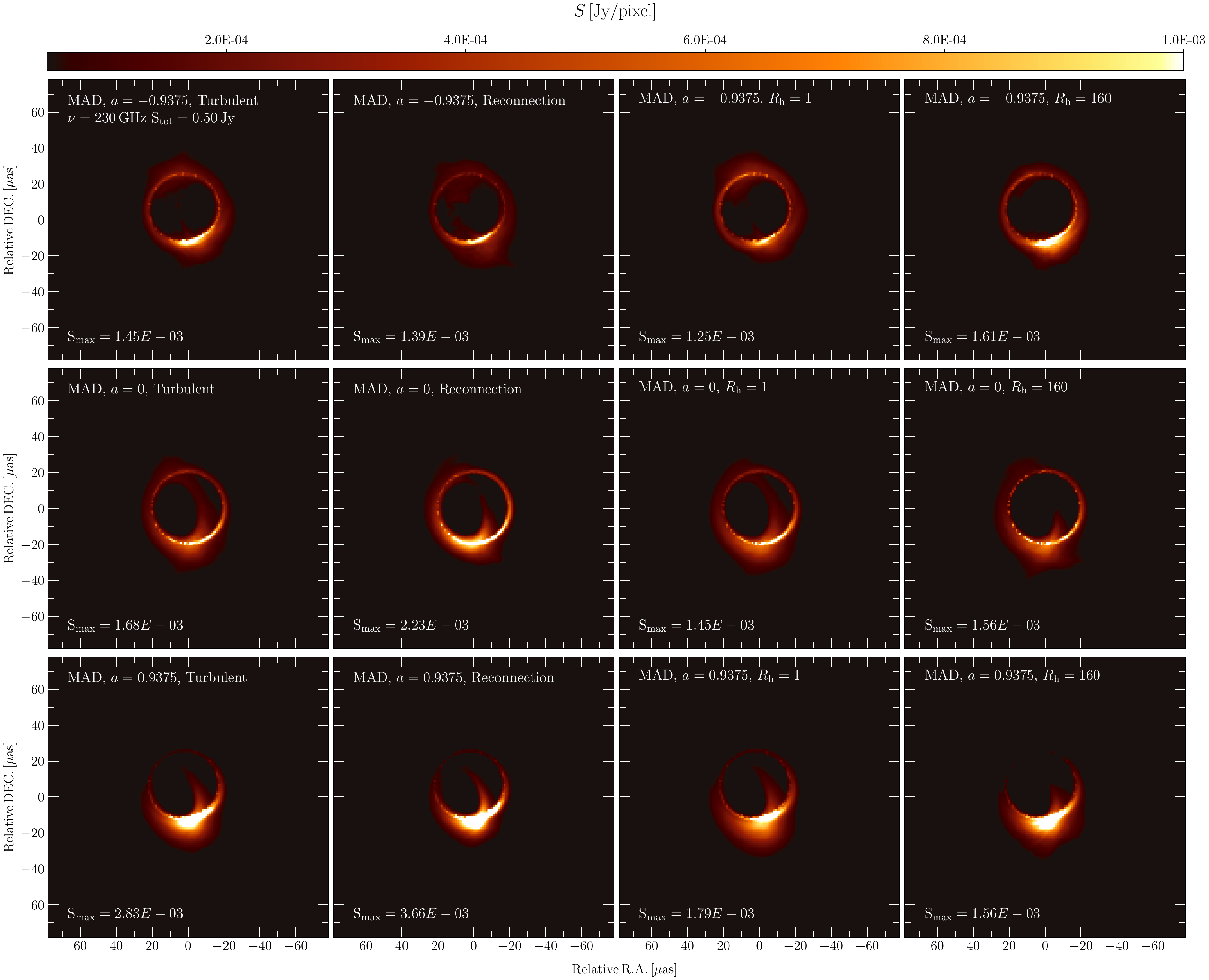}
 \caption{Same as Fig.~\ref{fig:GRRT_i163_lin}, but $i=60\degr$.}
 \label{fig:GRRT_i60_lin}
\end{center}
\end{figure*}

\begin{figure*}
\begin{center}
 \includegraphics[width=0.7\linewidth]{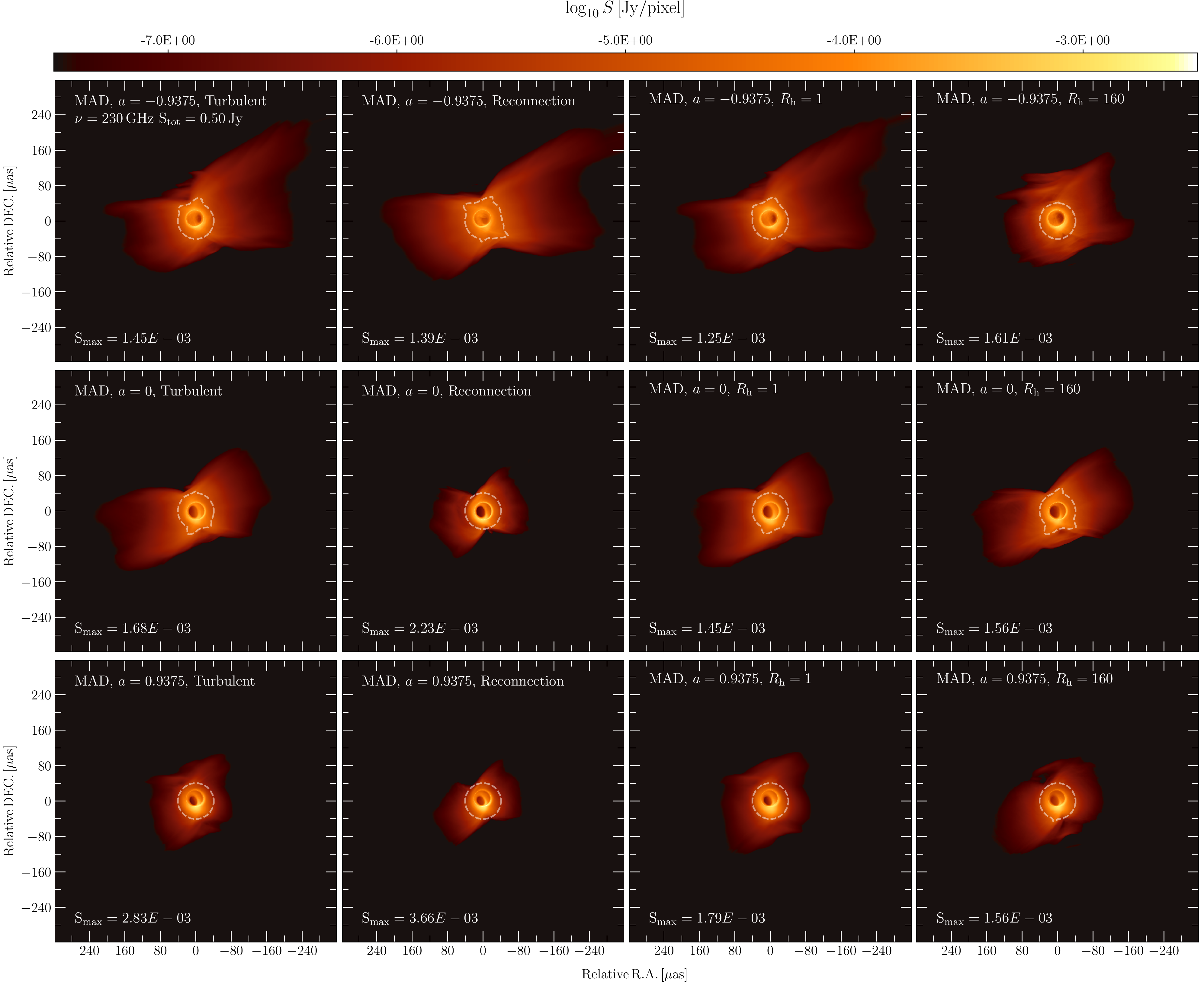}
 \caption{Same as Fig.~\ref{fig:GRRT_i163_log}, but $i=60\degr$.}
 \label{fig:GRRT_i60_log}
\end{center}
\end{figure*}

Figures~\ref{fig:GRRT_i60_lin} and \ref{fig:GRRT_i60_log} show the
time-averaged GRRT images of MAD simulations at $i=60\degr$ with
different black-hole spins in different ion-to-electron temperature ratio
prescriptions, in linear and logarithmic scales, respectively. At
$i=60\degr$, the asymmetry of the bright photon ring is more prominent
than the $i=163\degr$ case due to stronger Doppler beaming by the
emitting plasma. The general trends of different ion-to-electron ratio
prescriptions are unchanged. Images of turbulent heating prescriptions
have a more uniform distribution over the bright photon ring than other
models with lower maximum fluxes. In the logarithmic scale images, both
turbulent heating and magnetic-reconnection heating prescriptions have a
little widely extended diffused emission regions than the R-$\beta$
prescription, even though the R-$\beta$ prescription at $i=60\degr$ has a
more extended emission region than $i=163\degr$. Considering the
threshold of the 1\% of the maximum flux, this region is marginally
larger for the electron-heating model.

\subsection{Image Comparisons}
\begin{figure}
\begin{center}
 \includegraphics[width=1.0\linewidth]{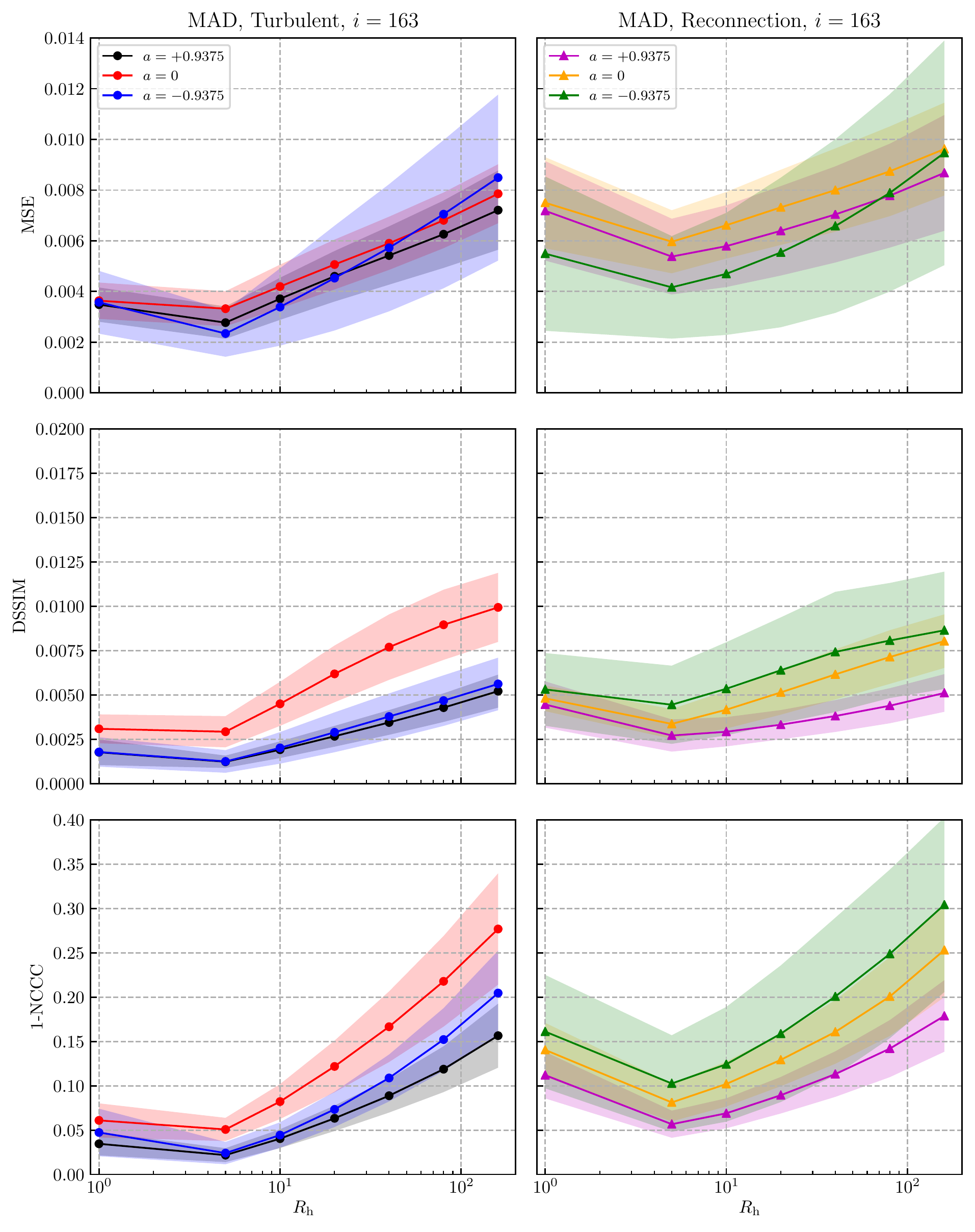}
 \caption{Image comparison distributions with different ion-to-electron
   ratio prescriptions using MSE (top), DSSIM (middle), and 1-NCCC
   (bottom). Left panels: comparison between turbulent heating and
   R-$\beta$ prescriptions with different black-hole spins, $a=0.9375$
   (black), $0$ (red), and $-0.9375$ (blue). Right panels: comparison
   between magnetic-reconnection heating and R-$\beta$ prescriptions with
   different black-hole spins, $a=0.9375$ (magenta), $0$ (orange), and
   $0.9375$ (green). Solid lines indicate the average value and same
   colour bands are the standard deviation of variation within each
   model. In all image-comparison metrics, a smaller value corresponds to
   a greater similarity between compared images.}
 \label{fig:comp-image_i163}
\end{center}
\end{figure}

In order to provide a more quantitative comparison between
electron-heating prescription models and R-$\beta$ models, we computed
three image-comparison metrics: the mean square error (MSE), structural
dissimilarity (DSSIM) \citep{Wang04}, and difference of normalised
cross-correlation coefficient (NCCC) from 1 \citep[see
  \eg][]{Mizuno2018,Fromm2020}.  The MSE is a pixel-by-pixel
  comparison metric calculated by averaging the squared intensity
  difference between two image pixels, namely
\begin{equation}
\mathrm{MSE}:= \frac{\sum^{N}_{j=1} |I_j - K_j|^2} {\sum^N_{j=1} |I_j|^2},
\end{equation}
where $I_j$ and $K_j$ are the $j$-th pixels of the images $I$ and $K$
with $N$ pixels.  The DSSIM is computed in terms of the human
visual-perception metric, also called the structural similarity index
(SSIM), so that $\mathrm{DSSIM} = 1/ |\mathrm{SSIM}| -1$. Given a
pair of images referred to as $I$ and $K$, the SSIM can be calculated as
\begin{equation}
\mathrm{SSIM (I, K)}:= \left( \frac{2 \mu_I \mu_K}{\mu_I^2 + \mu_K^2}\right) \left( \frac{2 \sigma_{IK}}{\sigma_I^2 + \sigma_K^2}\right),
\end{equation}
where $\mu_I := \sum^N_{j=1} I_j/N$, $\sigma^2_I=\sum^N_{j=1}
(I_j-\mu_j)^2/(N-1)$, and $\sigma_{IK} := \sum^N_{j=1} (I_j - \mu_I) (K_j
- \mu_K)/(N-1)$. For two images, $I$ and $K$, the NCCC is computed as
\begin{equation}
\mathrm{NCCC}:= \frac{1}{N} \sum_j \frac{(I_j - \langle I \rangle)(K_j - \langle K \rangle)}{\Delta_I \Delta_K},
\end{equation}
where $\langle I \rangle$ and $\langle K \rangle$ are the mean pixel
values in the images, and $\Delta_I$ and $\Delta_K$ are the standard
deviations of the pixel values in the two images
\citep[\eg][]{Akiyama2019_L4}. In other words, the NCCC quantifies the
similarity between two images, so that $\mathrm{NCCC}= 1$ corresponds to
a perfect correlation between the two images.

In these image comparisons, we use 101 snapshot GRRT images from
$t=14000\,M$ to $t=15000\,M$ (a $10M$ cadence) using both the
aforementioned electron-heating prescriptions and the R-$\beta$ model at
identical simulation times. We then check the variation of each
image-comparison metric. In all image-comparison metrics, smaller values
mean better matching in compared images. These image comparison results
for the 230\,GHz GRRT images at $i=163\degr$ are presented in
Fig. \ref{fig:comp-image_i163}. In general, as seen in
Fig.~\ref{fig:GRRT_i163_lin}, the R-$\beta$ prescription in the
  range of $R_\mathrm{h}=1$ to $160$ yields a good match for both
electron-heating prescriptions, as is reflected in the small values for
all three comparison metrics. As reference cases, we have performed two
additional comparisons. The first one is between a randomly chosen GRRT
image which is compared with each of the models considered. The second
case makes a comparison between the time-averaged GRRT images and the
individual snapshot GRRT images. Both results are summarised in
Appendix~\ref{sec:image_comparison_test}.

From the distribution of the image-comparison metrics, we do not see a
clear dependence of different black-hole spin cases in either comparison
with different electron-heating prescriptions, although specific average
values and the variance between models are different. For the
  turbulent-heating prescription, the cases with $R_\mathrm{h}=1$ and
  $R_\mathrm{h}=5$ have the smallest values of MSE, DSSIM and 1-NCCC.
Increasing the $R_\mathrm{h}$ value gives rise to an increase in the
image-comparison metrics.  For the magnetic-reconnection heating
prescription, the $R_\mathrm{h}=5$ case has the smallest value of MSE,
DSSIM and 1-NCCC.  From $R_\mathrm{h}=5$, both metrics become larger with
increasing $R_\mathrm{h}$ value. The smallest value of MSE ($\sim 0.004$)
and 1-NCCC ($\sim 0.05$) is more than a factor 2 smaller than the vales
of the comparison with a randomly chosen GRRT image (see
Fig.~\ref{fig:comp-image_i163_snap}). The smallest value of DSSIM is also
a factor of 2 smaller than that in the comparison with a randomly chosen
GRRT image in both electron-heating prescription (see
Fig.~\ref{fig:comp-image_i163_snap}). Note that in the majority of cases,
the comparison between different ion-to-electron ratio prescriptions
yields image metrics that are below those obtained from the comparison
with a randomly chosen or an average image.

However, variations for each case are large.  It is therefore difficult
to determine which cases are best matched with a given electron-heating
prescription on the basis of image-domain comparisons alone.  We
note that we have also applied the same image-domain comparison for the
$i=60$\degr case, obtaining similar results.

 The EHT observations of a black-hole shadow image are effectively
  limited by the finite angular resolution. To reproduced such a
  limitation, we have convolved the GRRT images with a 20 $\mu$as
  Gaussian beam and proceeded carrying out the same comparison discussed
  above. Obviously, the convolved GRRT images smear out the details of
  the images and, as a result, the difference between the images becomes
  intrinsically smaller. Hence, the values of all image-comparison
  metrics are reduced and less dependent on value chose for
  $R_\mathrm{h}$, although they still maintain the general trend seen
  in Fig.~\ref{fig:comp-image_i163}.

\begin{figure}
\begin{center}
 \includegraphics[width=1.0\linewidth]{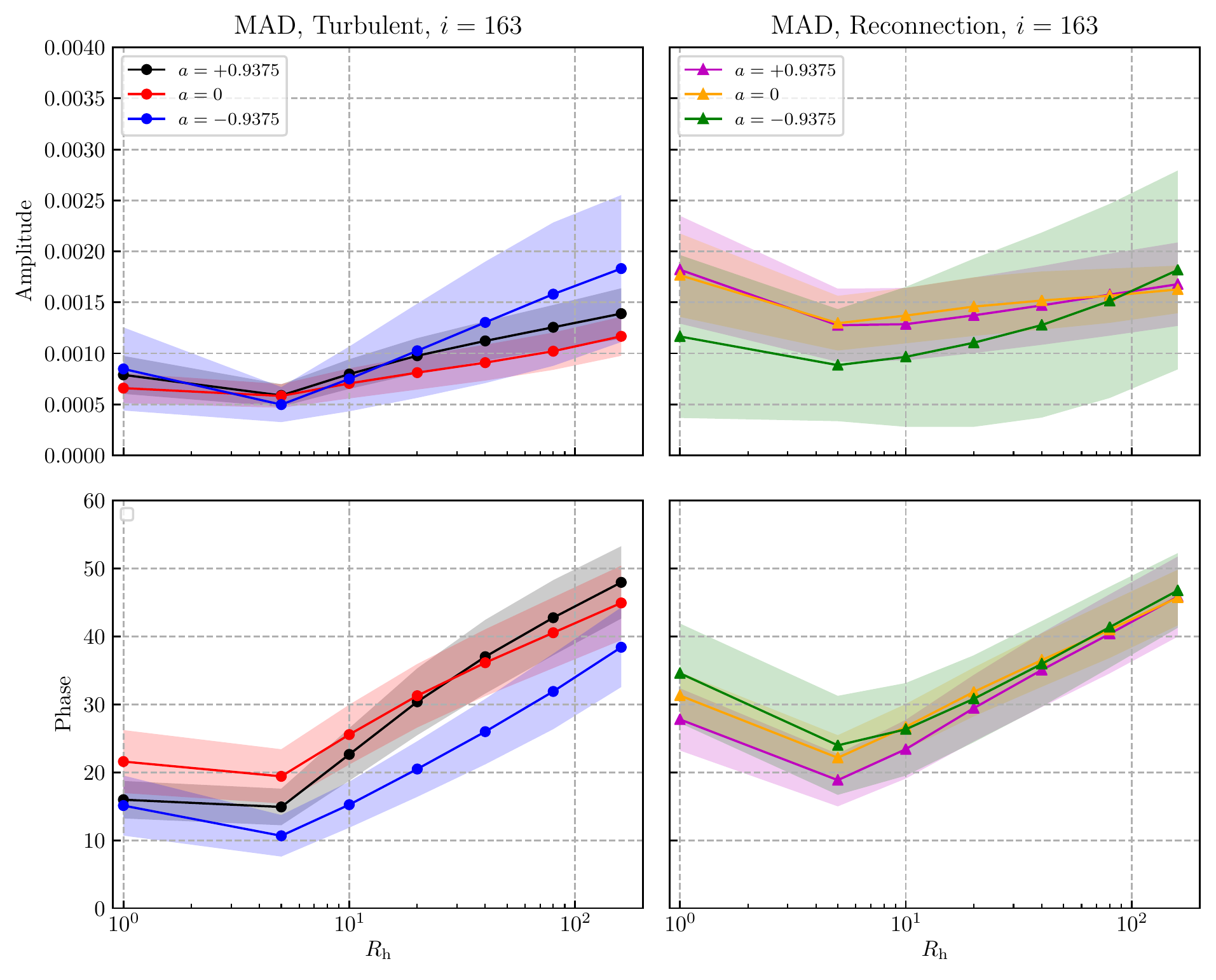}
\caption{Distribution of comparison of different ion-to-electron ratio
  prescriptions in visibility amplitude (top) and phase (bottom). Left
  panels: comparison between turbulent heating and R-$\beta$
  prescriptions with different black-hole spins, $a=0.9375$ (black), $0$
  (red), and $-0.9375$ (blue).  Right panels: comparison between
  magnetic-reconnection heating and R-$\beta$ prescriptions with
  different black-hole spins, $a=0.9375$ (magenta), $0$ (orange), and
  $-0.9375$ (green). Solid lines indicate average value and same colour
  bands indicate the standard deviation of each model.}
 \label{fig:comp-vis_i163}
\end{center}
\end{figure}

Additionally, we compare the different ion-to-electron ratio
prescriptions in the visibility domain. As for the image domain
comparison, we again use 101 snapshot GRRT images from $t=14000\,M$ to
$t=15000\,M$. For every image, we computed the Fourier transform,
  where we limited the baseline length to 10\,G$\lambda$\footnote{roughly
  the scale the longest EHT baselines probe}. From the Fourier transform,
  we compute the 2D distribution of the visibility amplitude (VA) and of
  the visibility phase (VP). Next, we calculate the mean VA and VP using
  all of our 101 Fourier-transformed GRRT snapshots. Finally, we compute
  the MSE between the mean and the individual VAs and VPs. Using this
  procedure, we do not limit our analysis to the specific baselines (and
  their orientation) of current observing arrays and thus we provide an
  array-independent comparison of our models in Fourier space.
Figure~\ref{fig:comp-vis_i163} shows the distribution of comparison
between turbulent heating and R-$\beta$ prescriptions (right panels) and
between magnetic-reconnection heating and R-$\beta$ prescriptions (left
panels) in terms of visibility amplitude and phases at $i=163$\degr.  The
general trends are the same as those seen in the image-domain comparison
(see Fig.~\ref{fig:comp-image_i163}).  A dependence on different
black-hole spin cases is again not present. In the turbulent heating
model, lower $R_\mathrm{h}$ cases have smaller differences in visibility
amplitude and phase. The differences in visibility amplitude and phase
become larger for larger $R_\mathrm{h}$ cases. In the magnetic
reconnection heating prescription, the $R_\mathrm{h}=5$ case has a
minimum in the differences of visibility amplitudes and phases.

From these quantitative comparisons of electron-heating prescriptions and
R-$\beta$ prescriptions in both the image and visibility domains, the
R-$\beta$ prescriptions are well matched to both the turbulent heating
and magnetic-reconnection heating models in 230\,GHz images. In general,
smaller $R_\mathrm{h}$ values are better matched to both heating
prescriptions, but variations within each model are large. It can
therefore not be said which model is better matched from the current
quantitative comparisons.

\subsection{Spectral Energy Distribution}

\begin{figure}
\begin{center}
 \includegraphics[width=1.0\linewidth]{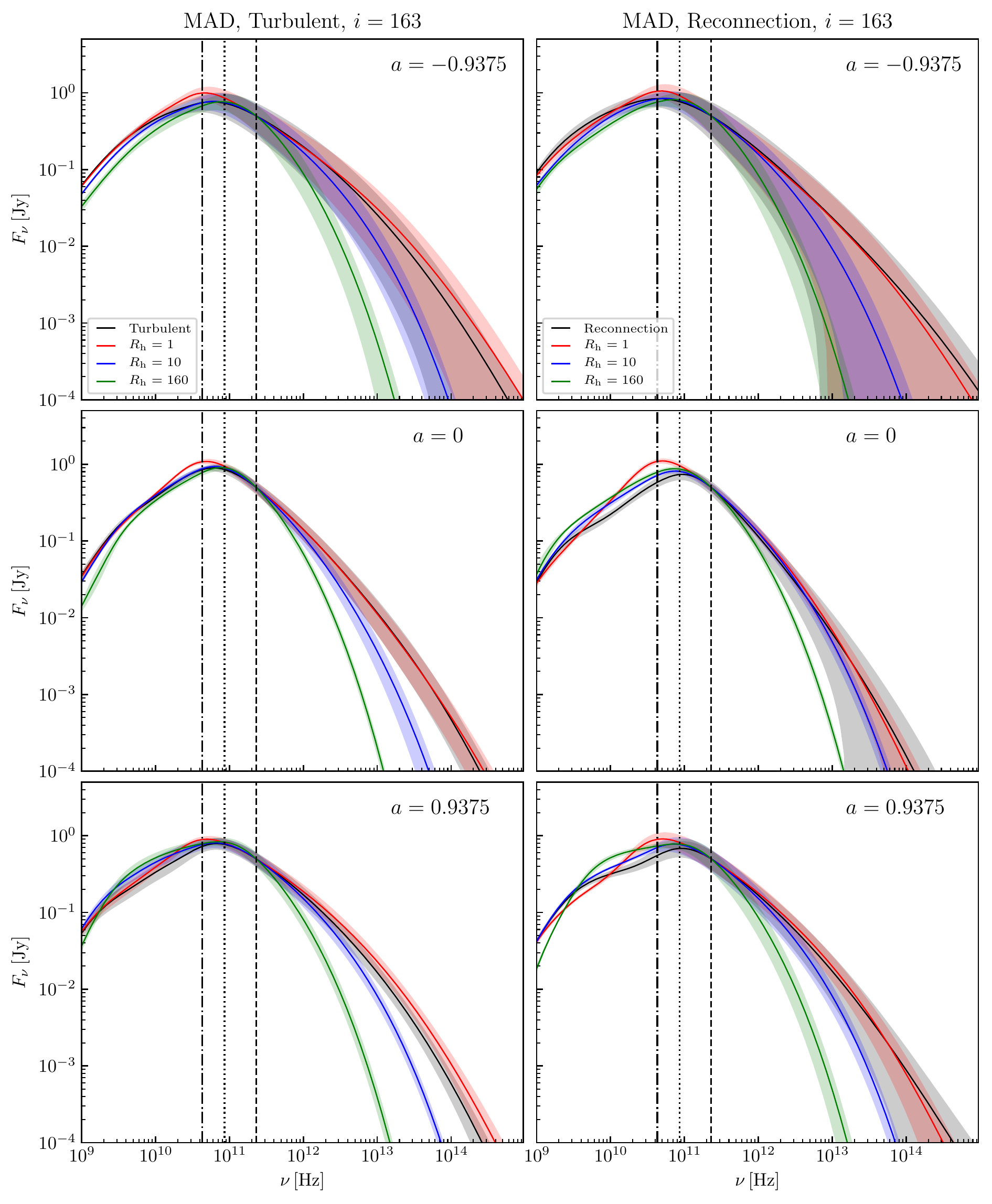}
 \caption{The spectral energy distributions (SEDs) of different
   ion-to-electron temperature ratio prescriptions: electron-heating
   model (black) and $R_\mathrm{h}=1$ (red), $10$ (blue), and $160$
   (green).  These are for R-$\beta$ model with different black-hole spin
   cases ({\it top}: $a=-0.9375$, {\it middle}: $0$, and {\it bottom}:
   $0.9375$) at $i=163\degr$. Solid lines indicate the average value and
   same colour bands denote the standard deviation of time variation of
   the spectrum. Vertical lines indicate frequencies of 43 (dash-dotted),
   86 (dotted), and 230\,GHz (dashed).}
 \label{fig:SED_i163}
\end{center}
\end{figure}

Broadband spectral energy distributions (SEDs) generated from different
ion-to-electron ratio prescriptions with different black-hole spin cases
at $i=163\degr$ are presented in Figure~\ref{fig:SED_i163}. In this
calculation, we fix the mass and distance to correspond to M87, and the
mass-accretion rate is fixed so that the average total flux at 230\,GHz
is 0.5\,Jy. Therefore all cases have the same averaged value at 230\,GHz
(vertical dashed line). In each model, solid lines indicates average
values and shaded regions denote the standard deviation of time
variation. In the R-$\beta$ prescription, smaller $R_\mathrm{h}$ value
cases have higher peak frequencies, shifted to lower frequencies. In both
lower and higher frequencies, the flux typically increases with lower
$R_\mathrm{h}$ values. This trend is generally seen in all black-hole
spin cases. Time variation of SEDs is smaller in higher $R_\mathrm{h}$
value cases, especially at higher frequencies. For the
  electron-heating prescription cases, both the turbulent and the
  reconnection-heating prescriptions behave similarly to the
  $R_\mathrm{h}=1$ case in the R-$\beta$ prescription at high
  frequencies, with the SED being peaked at 86\,GHz. Furthermore, when
  comparing with the R-$\beta$ model, the electron-heating prescriptions
  have a slightly higher flux at higher frequencies. This is reflected
in the fact that electrons are hotter than ions.
In the time variation of the SED, the
reconnection electron-heating prescription exhibits larger time
variations than turbulent electron heating at higher frequencies. From
these results, we conclude that electron-heating prescriptions yield
similar SED profiles to lower $R_\mathrm{h}$ values of the R-$\beta$
prescription at millimetre wavelengths. However, it will be possible to
distinguish using SEDs at higher wavelengths than the sub-mm.

\subsection{Image size}

\begin{figure}
\begin{center}
 \includegraphics[width=1.0\linewidth]{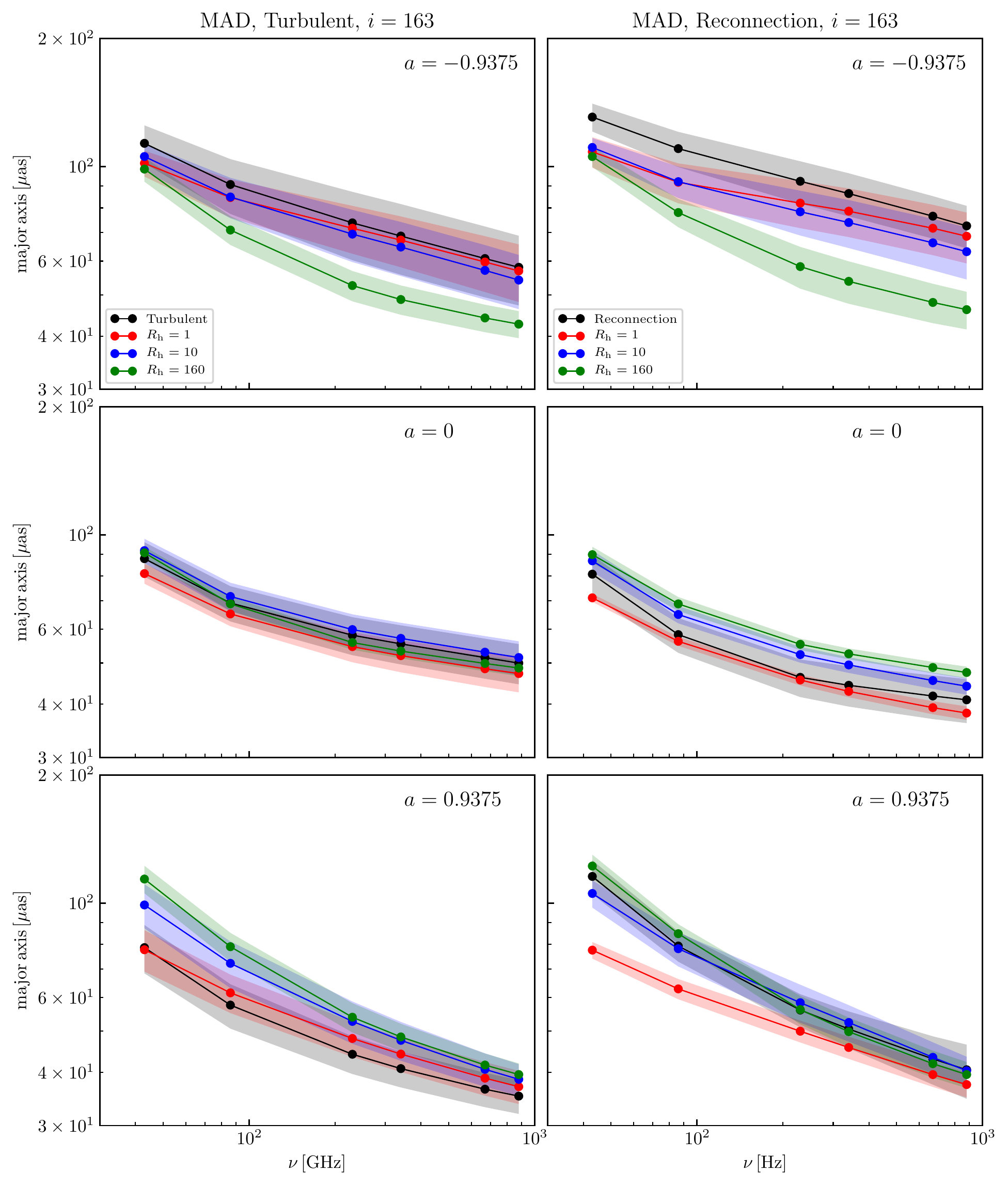}
 \caption{Image size (major axis) measured from the GRRT images at
   different frequencies (43\,GHz to 880\,GHz) using image moments in
   different ion-to-electron temperature ratio prescriptions with
   black-hole spin $a=-0.9375$ (top), $0$ (middle), and $0.9375$ (bottom)
   at $i=163\degr$. Different colour lines indicate different
   ion-to-electron ratio prescriptions: turbulent heating model (black),
   R-$\beta$ model with $R_\mathrm{h}=1$ (red), $10$ (blue), and $160$
   (green). Solid lines indicate the average value and same colour bands
   denote the standard deviation of time variation of the spectrum. }
 \label{fig:imagesize_i163}
\end{center}
\end{figure}

We compute the image size of the GRRT images at different frequencies
(43\,GHz to 880\,GHz) using image moments. The semi-major and semi-minor
axes of the images are computed from the eigenvalues $\lambda_{1,2}$ of
the covariance matrix formed by the second-order central-image
moments. The final FWHM of an equivalent elliptical Gaussian is computed
as $\Theta_{\rm major/minor}=\sqrt{8\ln(2)\lambda_{\rm 1,2}}$ \citep[see,
  \eg][]{Davelaar2019}. The result of this analysis can be found in
Fig. \ref{fig:imagesize_i163}. The different panels show the size of the
major axis of the ellipse enclosing the image structure for different
heating models (left: turbulent heating and right: reconnection heating)
for different black-hole spins (top: $a=-0.9375$, middle: $a=0$ and
bottom: $0.9375$). In each panel, the black curve indicate the heating
model and the other colours correspond to the R-$\beta$ model. In
  particular, for $a=-0.94$, the electron-heating prescriptions yield a
  larger image size, especially at high frequencies, due to the diffused
  extended emission seen in Fig. \ref{fig:GRRT_i163_log}. This result
also agrees well with the larger flux densities at higher frequencies
which is shown in the SED (see Fig \ref{fig:SED_i163}).

\subsection{Time variability}

\begin{figure}
\begin{center}
\includegraphics[width=1.0\linewidth]{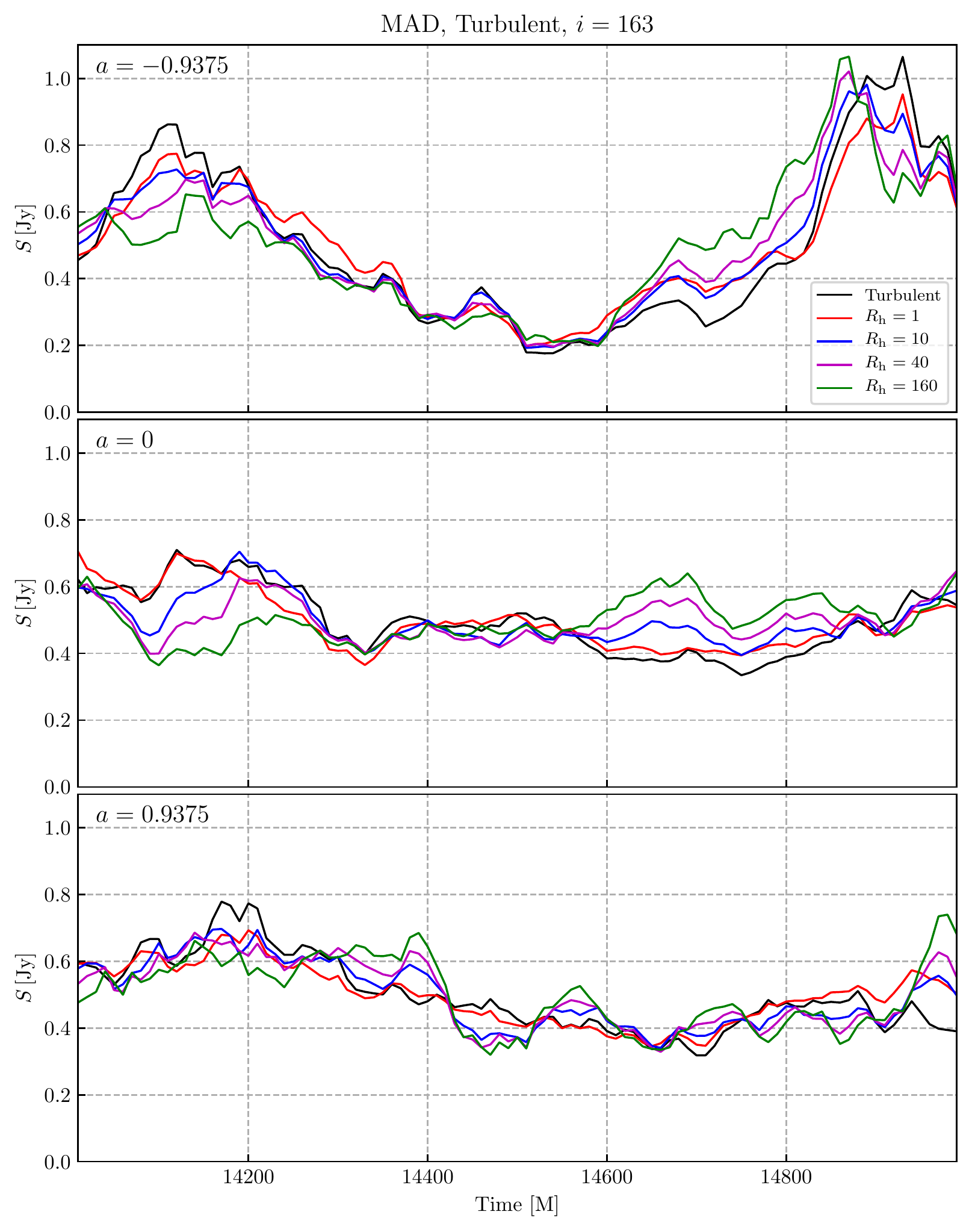}
\caption{Light curve of total flux at 230\,GHz in different
  ion-to-electron temperature ratio prescriptions with black-hole spin
  $a=-0.9375$ (top), $0$ (middle), and $0.9375$ (bottom) at
  $i=163\degr$. Different colour lines indicate different ion-to-electron
  ratio prescriptions: turbulent heating model (black), R-$\beta$ model
  with $R_\mathrm{h}=1$ (red), $10$ (blue), $40$ (magenta), and $160$
  (green).}
 \label{fig:Totflux_i163}
\end{center}
\end{figure}

\begin{figure*}
\begin{center}
 \includegraphics[width=0.85\linewidth]{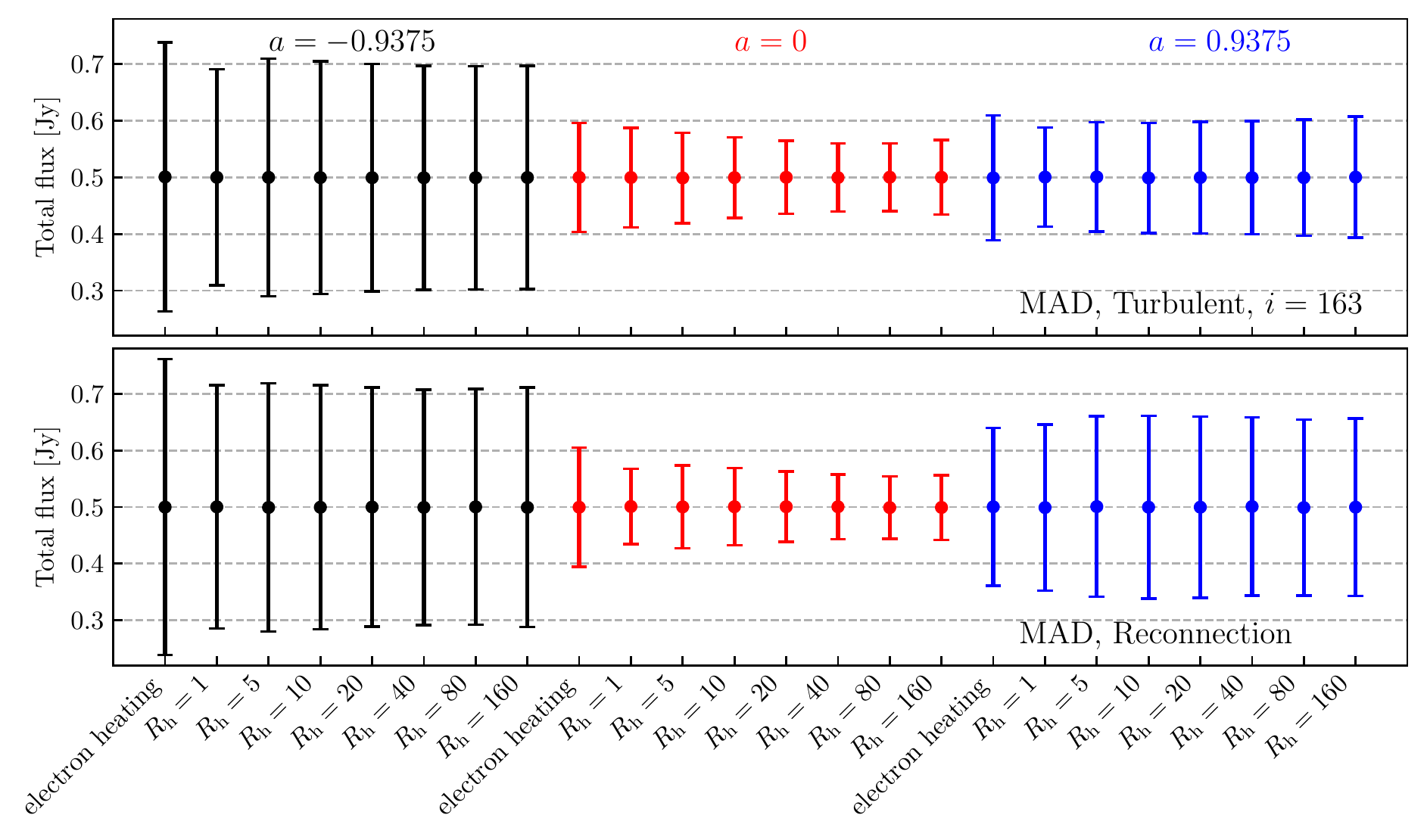}
 \caption{Fraction of total flux variation at 230\,GHz in different
   ion-to-electron temperature ratio prescriptions ({\it top}: turbulent
   heating model, {\it bottom}: reconnection heating model). Dotted point
   is averaged value and error bar is standard deviation of
   variance. Different colour indicates different black-hole spin cases,
   $a=-0.9375$ (red), $0$ (black), and $0.9375$ (blue)}
 \label{fig:Totalflux_avrg_i163}
\end{center}
\end{figure*}

Figure \ref{fig:Totflux_i163} shows the light curve of the total flux at
230\,GHz for different ion-to-electron ratio prescriptions with different
black-hole spin cases, at $i=163\degr$. Although we see some small
differences in variability at short timescales, all cases follow a
trend of large variations in the light curves. The quantitative
measurement of the fraction of total flux variation at 230\,GHz is shown
in Fig.~\ref{fig:Totalflux_avrg_i163}. In the comparison with different
ion-to-electron temperature ratio prescriptions, dependence on the
fraction of total flux variation at 230\,GHz is not seen. However, we see
some dependence in different black-hole spin cases. This shows that
retrograde spins have larger fractions of total flux variation at
230\,GHz than non-rotating and prograde spin cases. The non-rotating
black hole case has the smallest total flux variation. This may be a
consequence of the activity of orbiting flux tubes which are violent
episodes of flux escape from the black hole magnetosphere
\citep{Porth2020}, requiring a modestly spinning black hole. The
counter-rotating black hole case typically has a larger amount of
magnetic energy contained within orbiting flux tubes.

\subsection{Exclusion of Magnetised Region}

Due to the uncertainty of the results of the numerical simulations in the
highly magnetised region of the polar funnels, we have necessarily
excluded regions with $\sigma=1$ for the calculation of the GRRT
images. \cite{Chael2019} has investigated the dependence of the different
$\sigma$ thresholds on the 230\,GHz images and broadband spectra for M87
in two-temperature radiative GRMHD simulations of a MAD model. It was
found that incorporating radiative feedback from regions with $\sigma >
25$ makes the images more compact. For broadband spectra, the difference
in the images produced from different choices in the $\sigma$ threshold
is rather small, at least for frequencies up to 230\,GHz.

\begin{figure}
\begin{center}
 \includegraphics[width=1.0\linewidth]{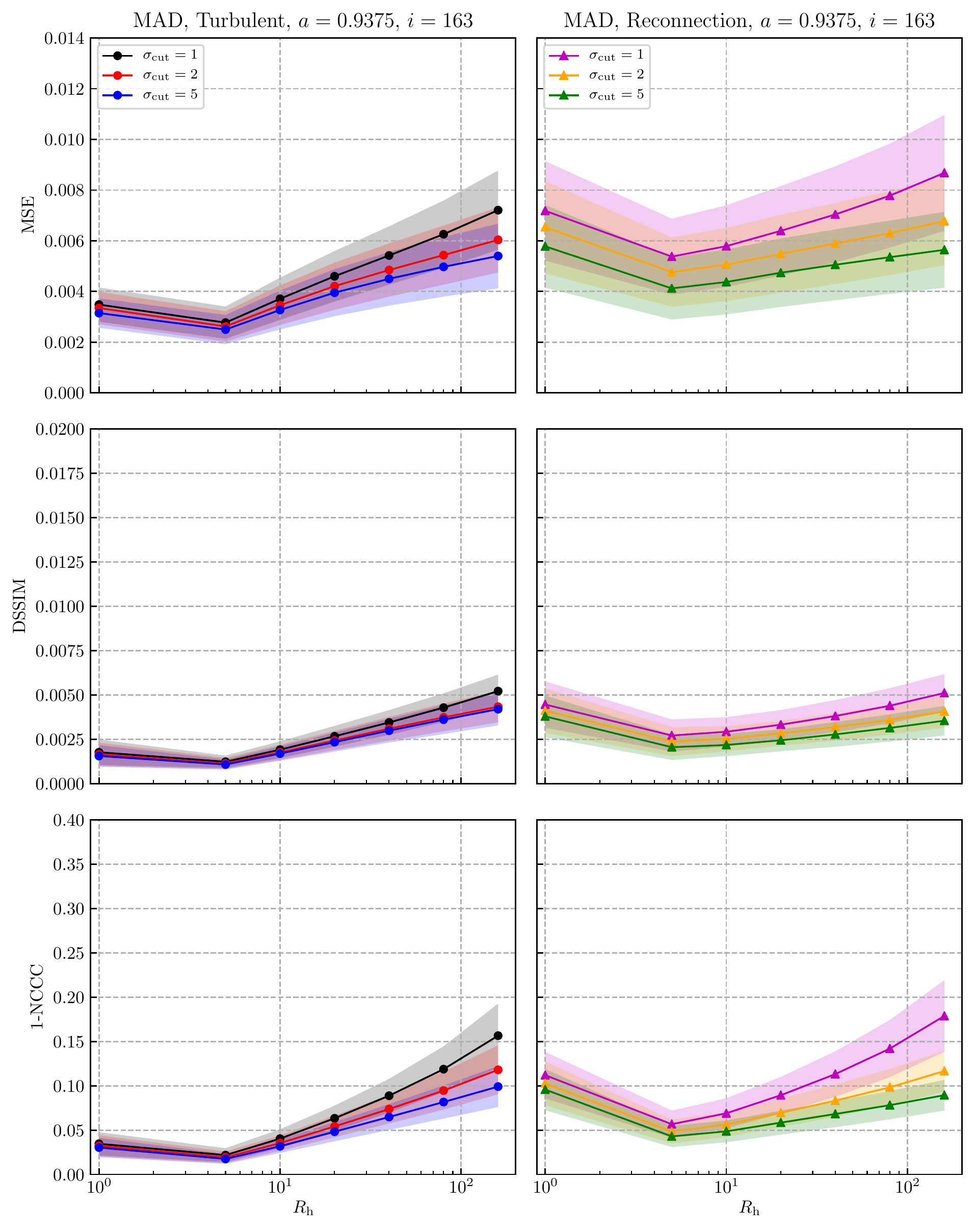}
 \caption{Same as Fig. \ref{fig:comp-image_i163} but different exclusion
   of magnetised region, $\sigma_\mathrm{cut}=1$ (black and magenta), $2$
   (red and orange), and $5$ (blue and green).}
 \label{fig:comp-image_i163_sigma}
\end{center}
\end{figure}
%
Here, we investigate the impact of the exclusion of different portions of
the highly magnetised regions on the image comparison. The results on the
comparison for the 230~GHz GRRT images at $i = 163$\degr using different
criteria for the exclusion of highly magnetised regions are presented in
Fig.~\ref{fig:comp-image_i163_sigma}. In essence, it is clear that
similar images are obtained when using different values of the cut-off
$\sigma$. For both the turbulent- and the magnetic-reconnection
  heating prescriptions, the case with $R_\mathrm{h} = 5$ has the
  smallest values of MSE, DSSIM and 1-NCCC, so that the image-comparison
  metrics increase with larger values of $R_\mathrm{h}$.
Increasing the value of the cut-off $\sigma$ leads to smaller
values of the image metrics, but the difference is very small. As a
result, we can conclude that choosing different sigma cut-off values such
as $\sigma=5$ does not significantly affect our results.

\section{Discussion}
\label{sec:discussion}

In this study, we have focused on 230\,GHz images with two different
inclination angles ($i=60\degr$ and $163\degr$). Although the emission
morphology is not much different when viewed in a linear scale, the
electron-heating prescriptions have a more extended and diffused emission
than the R-$\beta$ prescription. Because of the limited dynamical range
of the 2017 EHT observations, the images produced do not allow us
to distinguish the presence of and extend jet structure, thus preventing
from a clear identification of the electron-heating mechanisms at play in
M87*.  Such an extended emission component will be more clearly seen at
lower frequencies such as 43\,GHz and 86\,GHz
\citep[\eg][]{Chael2019,Davelaar2019,Chatterjee2020}. Therefore the
difference between electron heating and R-$\beta$ prescriptions may be
greater at longer wavelengths. Some indications of this have already been
seen in the broadband SEDs for different ion-to-electron temperature
ratio prescription values.  Near future observations 230\,GHz --
performed either by the next-generation EHT (ngEHT) or from space
\citep{Doeleman2019,Raymond2021,Roelofs2021} -- are expected to improve
the both the coverage in the $(u,v)$ plane and the dynamical range, thus
allowing for the detection of a diffused extended-jet structure and, in
turn, providing the evidence to distinguish the different potential
electron-heating mechanism.

Using a relativistic thermal electron distribution function is one of the
limitations of the present study, which will break down in regions where
non-ideal effects for magnetic fields are important. These non-ideal
effects are expected to be strong in highly magnetised regions such as
the jet funnel, where they can be associated with electron acceleration
mechanisms such as magnetic reconnection or
turbulence. \cite{Davelaar2018,Davelaar2019} use a $\kappa$ electron
distribution function, which is a combination of a thermal core with a
power-law tail at larger electron Lorentz factors, for GRMHD simulations
of M87. Although images at 230\,GHz do not show much difference,
non-thermal emission will produce more prominent and extended emission at
lower frequencies such as 43\,GHz and 86\,GHz or at a higher frequencies
as 1.1\,THz \citep{Petersen2020}. 
Distributions like the $\kappa$ electron distribution function
yield a better fit of the broadband spectrum of M87 \citep{Davelaar2019},
which we will investigate in the context of the present electron-heating
prescription for thermal core in a future work.

Our choice of accretion model, the MAD model, is the extreme limit of
high magnetic flux accretion onto a black hole horizon. Another typical
model is so-called Standard Accretion and Normal Evolution (SANE;
\cite{Narayan2012,Sadowski2013c}). As seen in \cite{Akiyama2019_L5}, GRRT
images of SANE models at 230\,GHz have a greater dependence on
$R_\mathrm{h}$ values than MAD models. Differences between
electron-heating prescriptions and R-$\beta$ models may be larger than
the current comparisons in MAD models and these considerations
  prevent us from making here more general statements that may be
  falsified under different accretion conditions. We will perform a
similar investigation of the comparison of different ion-to-electron
temperature ratio prescriptions to extended SANE models in an upcoming
paper.

In the theoretical comparison of the observations of M87* by the
\citet{Akiyama2019_L5}, the large majority of the GRMHD models produced
were considered compatible with the observed shadow image. At the same
time, a class of these models were rejected when three additional
constraints were imposed, namely: a consistent radiative efficiency, no
overproduction of X-ray emission, and jet power compatible with
large-scale radio observations. More specifically, the radiative
efficiency was calculated as $\epsilon \equiv L_\mathrm{bol}/\dot{M}c^2$,
where $L_\mathrm{bol}$ is the bolometric luminosity. If such an
efficiency was found to be larger than that of a thin, radiatively
efficient disc \citep{Novikov:1973}, \ie $\epsilon > 0.2$, then the model
was rejected. In this way, a number of MAD-accretion models with low
values of $R_\mathrm{h}$ were ruled-out because they were found to be
radiatively inconsistent. This result, however, was mostly due to the
very low mass-accretion rates of these models, which had
$\dot{M}/\dot{M}_\mathrm{Edd} \le 10^{-6}$ and thus resulted in being
radiatively inefficient. However, the mass-accretion rate in our
MAD-model simulations with a low $R_\mathrm{h}$ is around $10^{-5}\,
\dot{M}_\mathrm{Edd}$ (see Appendix~\ref{sec:mass_acc_rate}), i.e., one
order of magnitude greater than that found by the
\citet{Akiyama2019_L5}. As a result, the corresponding radiative
efficiency of all of our MAD models -- even those with a low
$R_\mathrm{h}$ -- is still compatible with the constraints from a thin
disc, and hence they can all be considered compatible with the
observations. There are a number of potential sources for the different
mass-accretion rates measured both here and those considered by
\citet{Akiyama2019_L5}, but the different adiabatic index ($\Gamma_{\rm
  g}=4/3$ versus $\Gamma_{\rm g}=13/9$ in the previous work) is likely
the element most responsible for this difference.

We also note that a value of $\Gamma_{\rm g}=4/3$ reflects the assumption
that the plasma is relativistic. The initial gas torus might be cooler
and have a higher adiabatic index, \eg $\Gamma_g \simeq 5/3$. As seen in
Appendix~\ref{sec:shock_test}, differences in adiabatic index between gas
and electrons affects the efficiency of the electron heating. This effect
is seen in the dimensionless electron temperature distribution seen in
Fig.~\ref{fig:GRMHD_Thetae} and in previous studies
\citep{Chael2019,Dexter2020} However, the GRRT images we calculate have
the same general trend as those seen in \citet{Chael2019}. We therefore
conclude that our results are not affected appreciably by our choice of
adiabatic index, but they may be affected by the measurement of the
mass-accretion rate.

\section{Conclusions}
\label{sec:conclusion}

In this paper, we have investigated the commonly used ion-to-electron
temperature ratio prescription, the R-$\beta$ model, by computing GRRT
images at 230\,GHz which serve to facilitate comparison between
electron-heating prescriptions obtained from GRMHD simulations with
electron thermodynamics. From the comparison of GRRT images, the
R-$\beta$ prescription in the range of $R_\mathrm{h}$ from 1 to 160 with
fixed $R_\mathrm{l}=1$ is well-matched by both turbulent heating and
magnetic-reconnection heating prescriptions, although images of
electron-heating prescriptions have a more extended and diffused emission
region. From this comparison study of different physical aspects,
including images, visibilities, broadband spectra, and light curves, we
conclude that the commonly-used R-$\beta$ model favourably reproduces the
ion-to-electron temperature prescription obtained from electron
thermodynamics calculations of accretion flows onto a black hole at
230\,GHz. In general, smaller $R_\mathrm{h}$ values yield a better match
to both heating prescriptions. For observations at longer
  wavelengths, such as 43\,GHz or 86\,GHz, a greater difference in the
  images due to the more extended and diffused emission in the
  electron-heating prescription cases is expected, in particular for the
  counter-rotating accretion. We note that our conclusions apply to MAD
accretion models, and it is expected that greater differences will be
found for SANE models, which exhibit much greater variability than MAD
models.

\section*{Acknowledgements}
We would like to thank Alejandro Cruz Osorio, Antonios Nathanail, Jonas
K\"{o}hler, Roman Gold, Mariafelicia de Laurentis, and Avery Broderick,
for useful discussions. This research is supported by the ERC synergy
grant ''BlackHoleCam: Imaging the Event Horizon of Black Holes'' (grant
number 610058). CMF is supported by the Black Hole Initiative at Harvard 
University, which is supported by a grant from the John Templeton Foundation. 
ZY is supported by a Leverhulme Trust Early Career
Fellowship. The simulations were performed on GOETHE at the CSC-Frankfurt,
 Iboga at ITP Frankfurt, and Pi2.0 at Shanghai Jiao Tong University. 
This research has made use of NASA's astrophysics data system (ADS).

\section*{Data availability}
The data underlying this article will be shared on reasonable request to
the corresponding author.




\bibliographystyle{mnras}

\appendix
\section{1D Noh Shock Test}
\label{sec:shock_test}

\begin{figure*}
\begin{center}
 \includegraphics[width=0.7\linewidth]{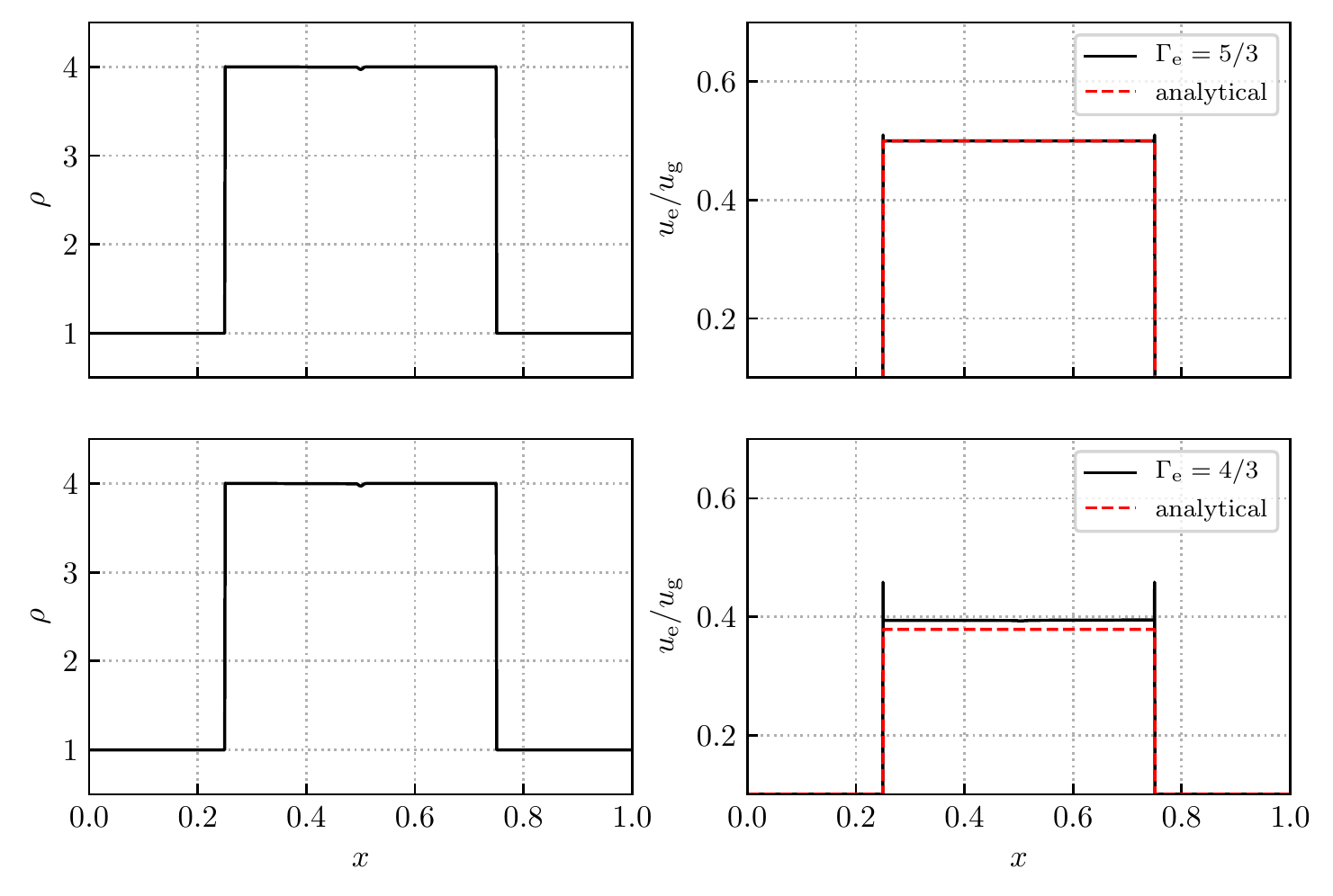}
 \caption{1D Noh shock results for an electron heating fraction $f_{\rm
     e} = 0.5$ using 2000 cells with ({\it top}) $\Gamma_{\rm e}=5/3$ and
   ({\it bottom}) $\Gamma_{\rm e}=4/3$. Left panels show density, right
   panels present $u_{\rm e}/u_{\rm g}$. Red dashed lines indicate
   analytical solutions. }
 \label{fig:1DNoh_shock}
\end{center}
\end{figure*}
%
We demonstrate the validity and convergence properties of our
implementation of the electron-heating prescription using a 1D Noh shock
test problem. The Noh shock test examines the problem of shock
reflection. The time-dependent solution consists of two shocks
originating at the initial discontinuity and travelling to the left and
right boundaries.

In high Mach number shocks, the electrons receive a constant fraction of
the viscous heating by the shock. As seen in \cite{Ressler2015}, the
post-shock electron internal energy $u_e^f$ is given by
\begin{equation}
\frac{u_{\rm e}^f}{u_{\rm g}^f} = \frac{f_{\rm e}}{2} \left[ \left( \frac{\Gamma_{\rm g}+1}{\Gamma_{\rm g}-1} \right)^{\Gamma_{\rm e}} \left( 1- \frac{\Gamma_{\rm g}}{\Gamma_{\rm e}}\right) +1 + \frac{\Gamma_{\rm g}}{\Gamma_{\rm e}} \right] \frac{\Gamma_{\rm g}^2-1}{\Gamma_{\rm e}^2-1} \,,
\end{equation}
where $u_{\rm g}^f$ is post-shock internal energy of the fluid. When
$\Gamma_{\rm g} = \Gamma_{\rm e}$, $u_{\rm e}^f/u_{\rm g}^f$ is equal to
$f_{\rm e}$. When we choose $\Gamma_{\rm g}=5/3$ and $\Gamma_{\rm
  e}=4/3$, $u_{\rm e}^f/u_{\rm g}^f$ becomes $\sim 0.76 f_{\rm e}$.

In our simulations, we assume an unmagnetised, non-relativistic
($\Gamma_{\rm g}=5/3$) cold fluid with uniform density and gas pressure
as an initial condition. The initial velocities have discontinuities at
the half of computational domain (left and right states) and are directed
towards the discontinuous boundary with non-relativistic speed $|v|
  = 0.001c$. This creates a strong shock ($M \gg 1$) propagating in the
left and right directions. In this test, $f_{\rm e}$ is fixed as 0.5 and
the test is performed for both $\Gamma_{\rm e} = 5/3$ and $\Gamma_{\rm e}
= 4/3$.

\begin{figure}
\begin{center}
 \includegraphics[width=0.8\linewidth]{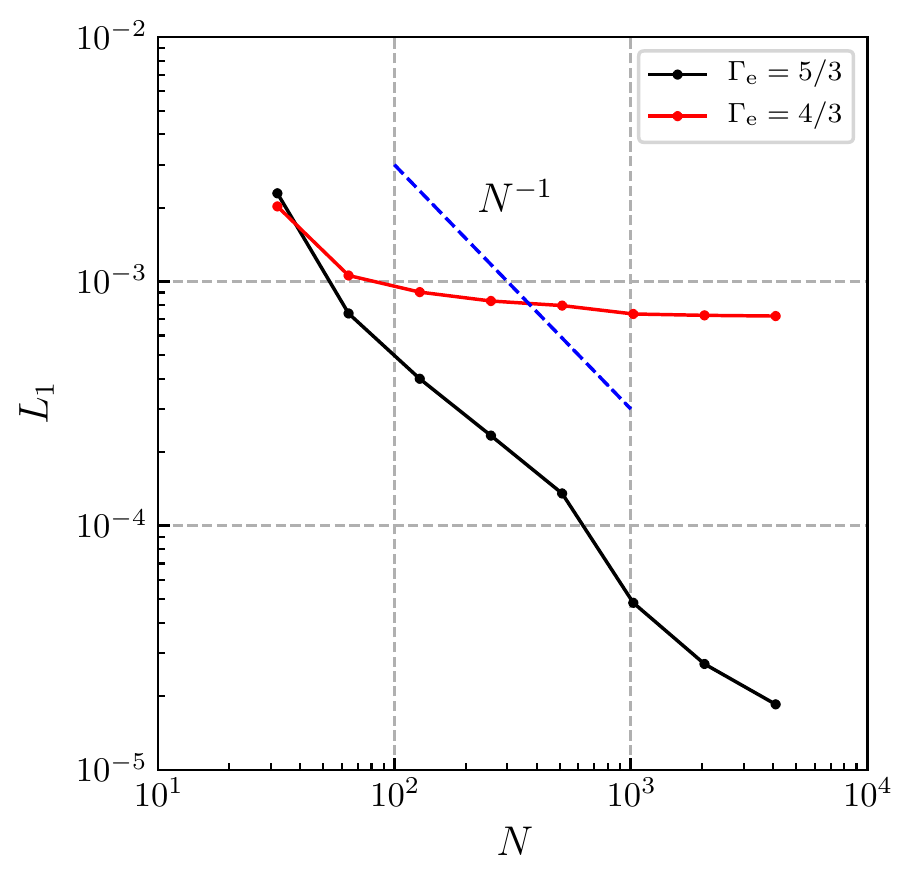}
 \caption{L1 norm of the error in the ratio of the electron-to-gas
   internal energies between the numerical and the analytical solutions
   shown as a function of the number $N$ of cells used.}
 \label{fig:1DNoh_shock_conv}
\end{center}
\end{figure}

Figure~\ref{fig:1DNoh_shock} shows the distribution of density and the
ratio of internal energy between electrons and gas at $t=700$. Two
shocks propagate at $x \approx 0.25$ and $0.75$. After the shock, the
density jumps to a value four times larger than the pre-shock region,
which indicates a strong shock has been created. In the $\Gamma_{\rm
  e}=5/3$ case, the ratio of internal energy between electrons and gas at
the post-shock region is in good agreement with the analytical
solution. However, for the $\Gamma_{\rm e}=4/3$ case there is not as
close a match with the analytical solution and differs by $\sim$
4\%. Similar results are also seen in \cite{Ressler2015}. This is because
the correct heating by the shock needs the shock structure to be
well-resolved. In our numerical scheme, the shock structure is still
resolved with a few grid points. We therefore cannot resolve the shock
completely, even at higher grid resolutions than those adopted in this
study. For the $\Gamma_{\rm g}=\Gamma_{\rm e}$ case, the density term is
cancelled out in the time evolution of the electron entropy, and its
dependence on the shock structure may be neglected.

The simulation convergence for the electron heating calculations is shown
in Fig.~\ref{fig:1DNoh_shock_conv}. As expected, the $\Gamma_{\rm e}=5/3$
case converges at first order (\ie as $1/N$, where $N$ is the number of
cells used), but the $\Gamma_{\rm e}=4/3$ case does not converge to the
analytical solution, as seen in Fig.~\ref{fig:1DNoh_shock} (numerical
results differ by $\sim 4\%$). In order to reach the analytical solution
in the $\Gamma_{\rm g} \neq \Gamma_{\rm e}$ case, we would need to
introduce dissipative effects and in particular a bulk viscosity
\cite{Ressler2015}.  Given that we are concerned primarily with synthetic
230\,GHz images, a $\le$ 4\% error is acceptable for the present study.

\section{Image Comparison Test}
\label{sec:image_comparison_test}

\begin{figure*}
\begin{center}
 \includegraphics[width=1.0\linewidth]{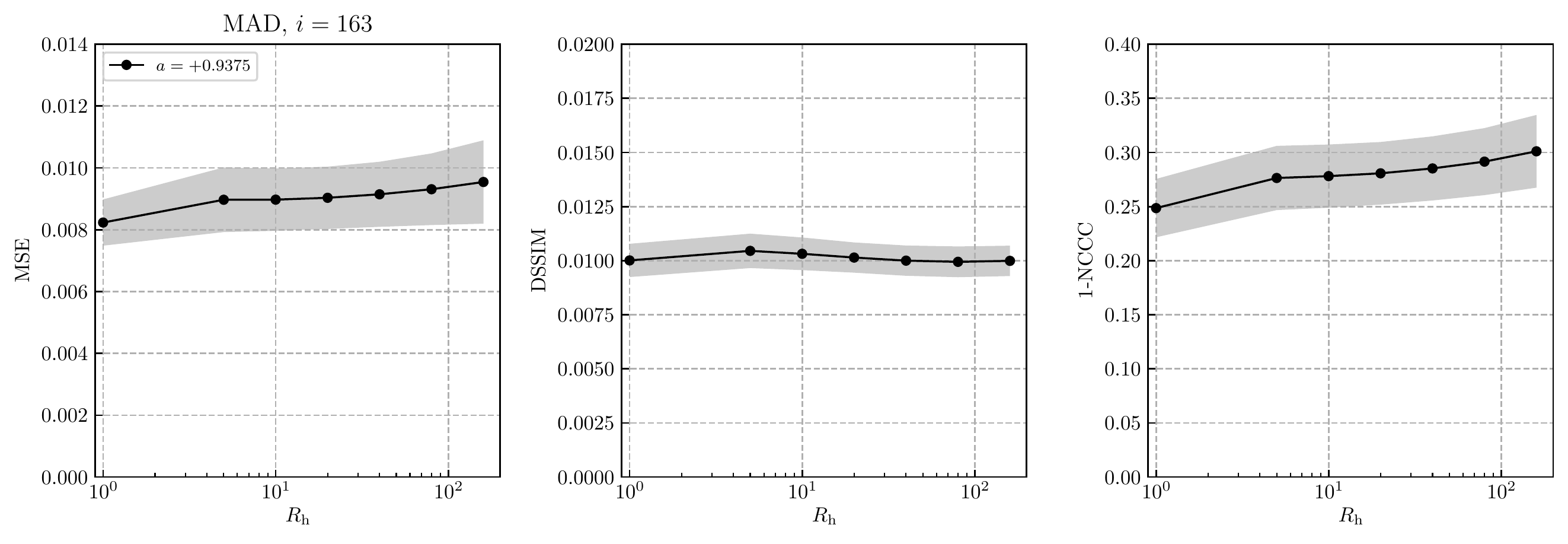}
 \caption{ Image-comparison distributions at an inclination angle
   $i=163$\degr for different values of $R_\mathrm{h}$ of the R-$\beta$
   prescription for a black hole with spin $a=0.9375$ and a single image
   relative to a black hole with spin $a=0$ with $R_\mathrm{h}=40$ at
   $t=15000\,M$. The different panels refer to the different metrics: MSE
   (right), DSSIM (middle), and 1-NCCC (left). Solid lines indicate the
   average values and the shaded bands are the standard deviation of the
   variation within each model.}
 \label{fig:comp-image_i163_snap}
\end{center}
\end{figure*}
%
In order to provide a reference value of the image-comparison metrics, we
consider the comparison between each prescription model ( different
  $R_\mathrm{h}$ value of $R-\beta$ parameterised prescriptions) and a
randomly chosen single image. For the latter, we consider the 230\,GHz
GRRT image of a black hole with spin $a=0$ using the $R-\beta$
parameterised prescription with $R_\mathrm{h}=40$ at $t=15000\,M$.

Figure \ref{fig:comp-image_i163_snap} shows the results of the image
comparison for the 230\,GHz GRRT images at $i=163$\degr with a black-hole
spin $a=0.9375$. From the distribution of the image-comparison metrics,
no evidence emerges for a clear dependence on the different heating
prescriptions. In other words, all models match the chosen reference
image equally poorly or favourably. The average values of the metrics
MSE, DSSIM, and 1-NCCC are $0.08-0.09$, $0.01$, and $0.25-0.3$,
respectively. Note that the MSE has a slightly larger variance for larger
values of $R_\mathrm{h}$. More importantly, these average values are
around a factor of two larger than the smallest value in
Fig.~\ref{fig:comp-image_i163}. We also note that the correlation time of
images in GRMHD simulations has been found to be around $50\,M$, that is,
any two random images of a GRMHD simulation do not show a correlation if
the corresponding time separation is larger than $50\,M$. Beyond this
window in time, the differences can be very significant.

%
\begin{figure*}
\begin{center}
 \includegraphics[width=1.0\linewidth]{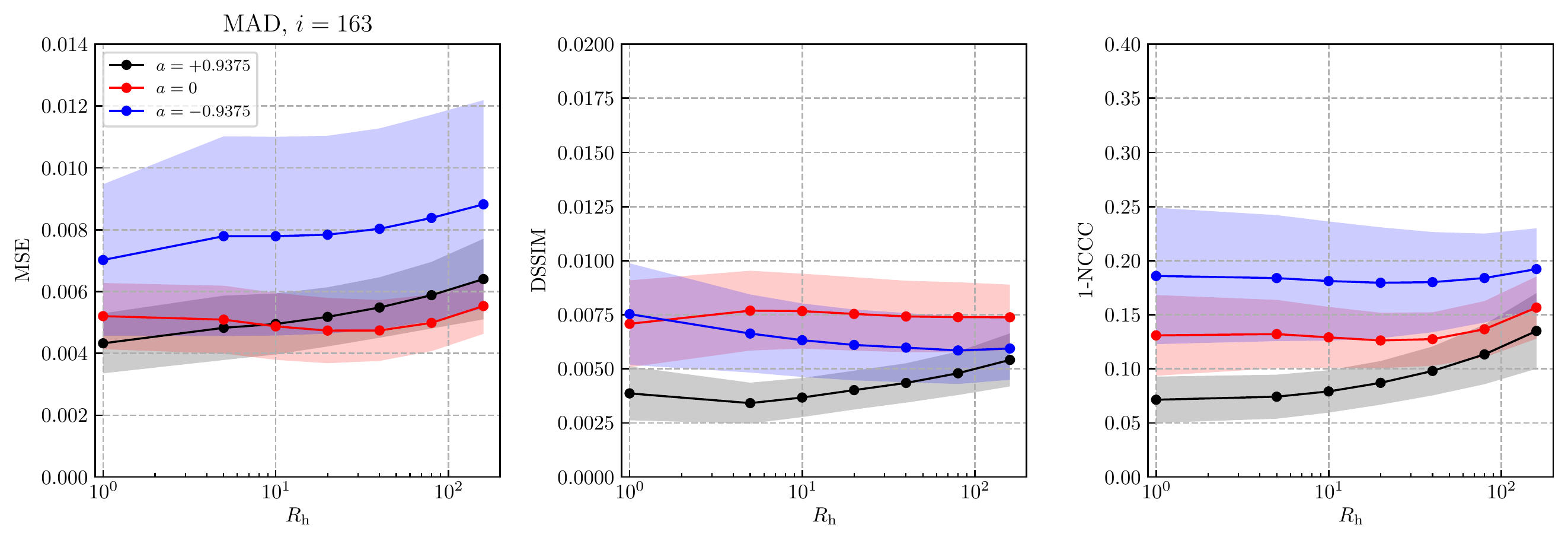}
 \caption{ Same as in Fig. \ref{fig:comp-image_i163_snap} but with the
   addition of the comparison with averaged images for black holes with different spins: blue
   for $a=-0.9375$, red for $a=0$, and black for $a=0.9375$.}
 \label{fig:comp-image_i163_average}
\end{center}
\end{figure*}
%
Another reference value of the image-comparison metrics is obtained by
the comparison between time-averaged GRRT images and individual snapshot
GRRT images in each prescription model with different black hole
  spin. This is shown in Fig.~\ref{fig:comp-image_i163_average}. Note
that all distributions are essentially flat, indicating a very weak
dependence on the value of $R_\mathrm{h}$. When $R_\mathrm{h}$ increases,
the values of the MSE and 1-NCCC metrics also increase, although only
slightly. Note also that the distribution of the MSE metric has a large
variance in the case of counter-rotating black holes, indicating that
each individual image can have a large variation from the averaged
one. This behaviour was already encountered in the fraction of the
total-flux variation in Fig.~\ref{fig:Totalflux_avrg_i163}.

\section{Different $R_\mathrm{l}$ case}
\label{sec:diff_Rlow}

\begin{figure*}
\begin{center}
 \includegraphics[width=0.9\linewidth]{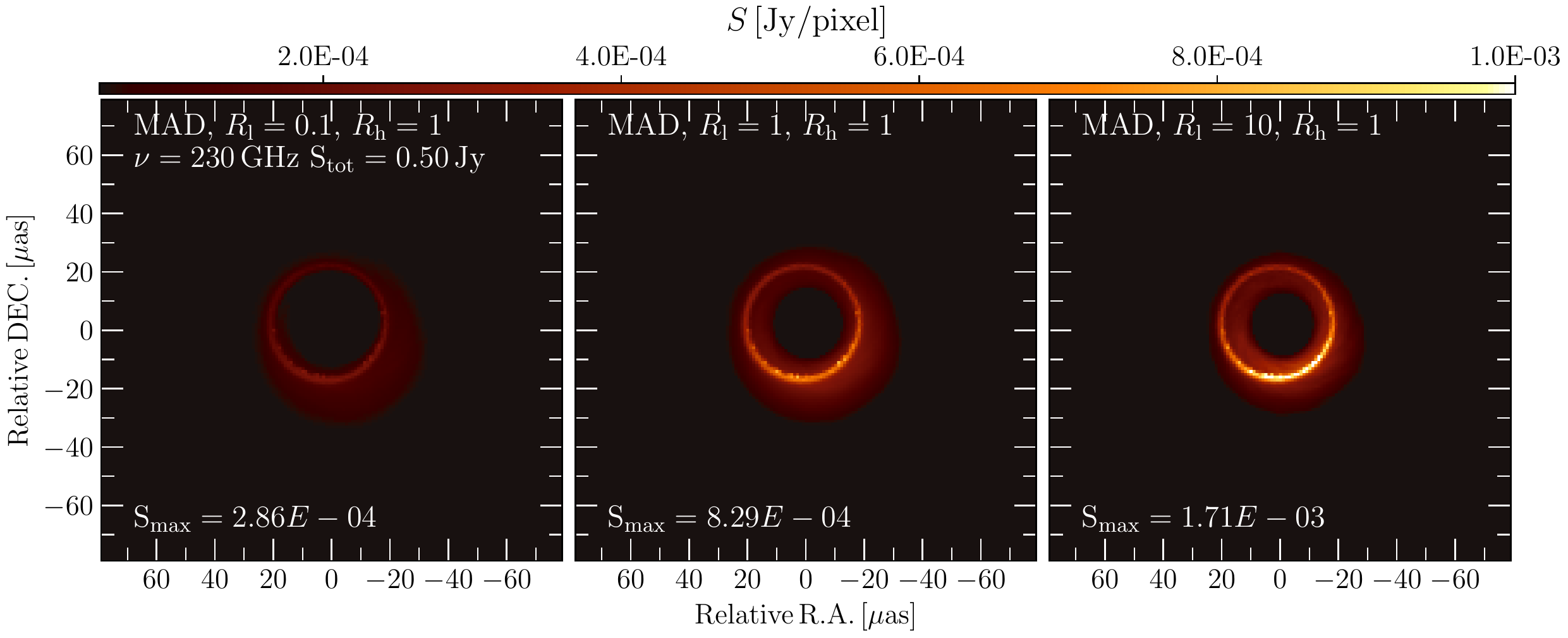}
 \includegraphics[width=0.9\linewidth]{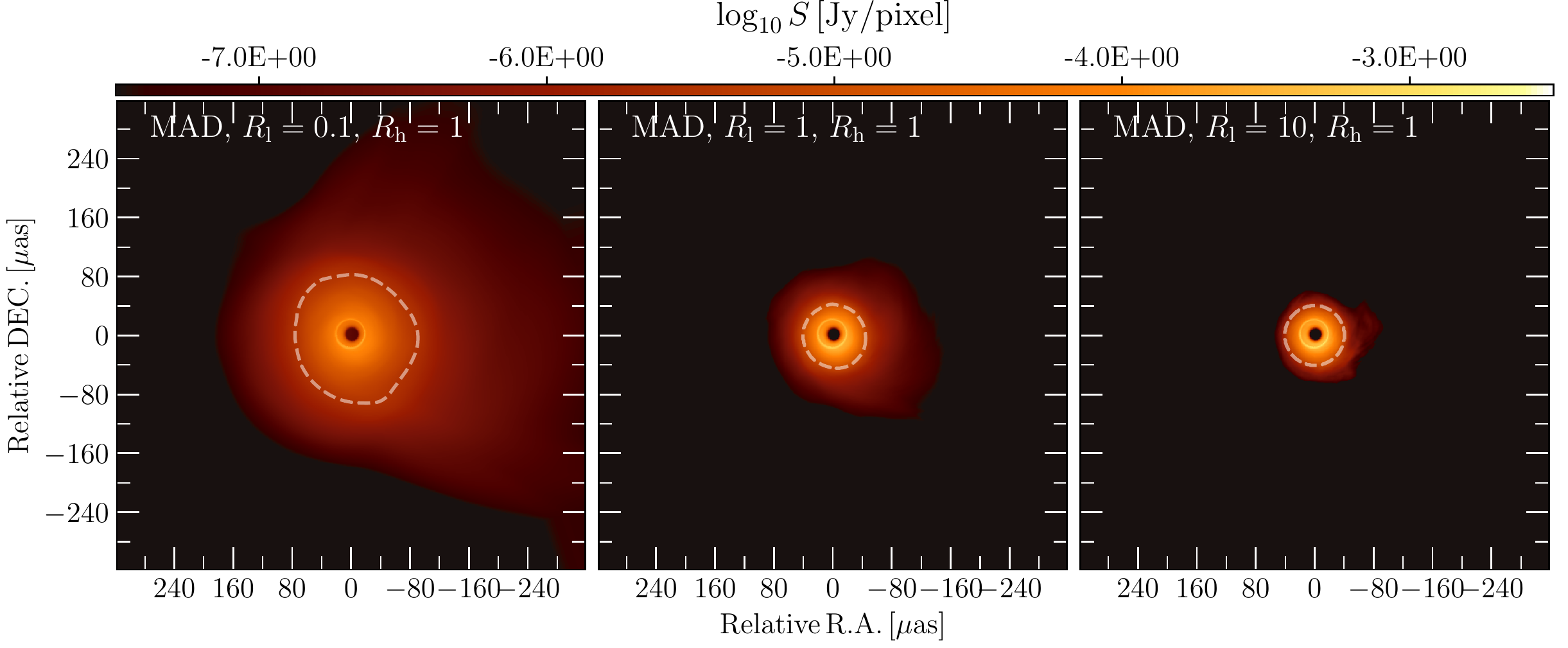}
 \caption{Time-averaged GRRT images at $i=163\degr$ of MAD simulations
   with a black hole spin $a=0.9375$ using the $R-\beta$ prescription
   with $R_\mathrm{h}=1$ and $R_\mathrm{l}=0.1$, 1, and 10.  Upper and
   lower panels show linear and logarithmic scale respectively. The image
   is averaged with GRRT images from $t = 14000\,M$ to $15000\,M$. All
   averaged images have the same total flux with 0.5\,Jy at 230 GHz.  }
 \label{fig:GRRT_i163_Rl}
\end{center}
\end{figure*}

We recall that the parameterised R-$\beta$ prescription for the
  electron-ion temperature ratio has two parameters, $R_\mathrm{l}$ and
  $R_\mathrm{h}$. As mentioned in the main text, we have kept
  $R_\mathrm{l}=1$ fixed and varied $R_\mathrm{h}$. In this appendix, we
  investigate the effect of actually using different values of
  $R_\mathrm{l}$. In particular, Fig.~\ref{fig:GRRT_i163_Rl} presents the
  time-averaged GRRT images at $i=163\degr$ relative to MAD simulations
  with a black hole spin $a=0.9375$ using the $R-\beta$ prescription with
  $R_\mathrm{h} =1$ and $R_\mathrm{l}=0.1$, $1$, and $10$, both in a
  linear and in a logarithmic scale. Following
  Figs.~\ref{fig:GRRT_i163_lin} and \ref{fig:GRRT_i163_log}, the images are
  averaged from $t=14000\,M$ to $15000\,M$, and all averaged images have
  the same total flux of 0.5\,Jy.  Note that for values of $R_\mathrm{l}$
  the bright photon ring becomes dimmer and a larger amount of an
  extended and diffused emission is seen. On other hand, for large values
  of $R_\mathrm{l}$ the emission is more concentrated near the photon
  ring and the extended emission less pronounced.

%

\begin{figure*}
\begin{center}
 \includegraphics[width=0.9\linewidth]{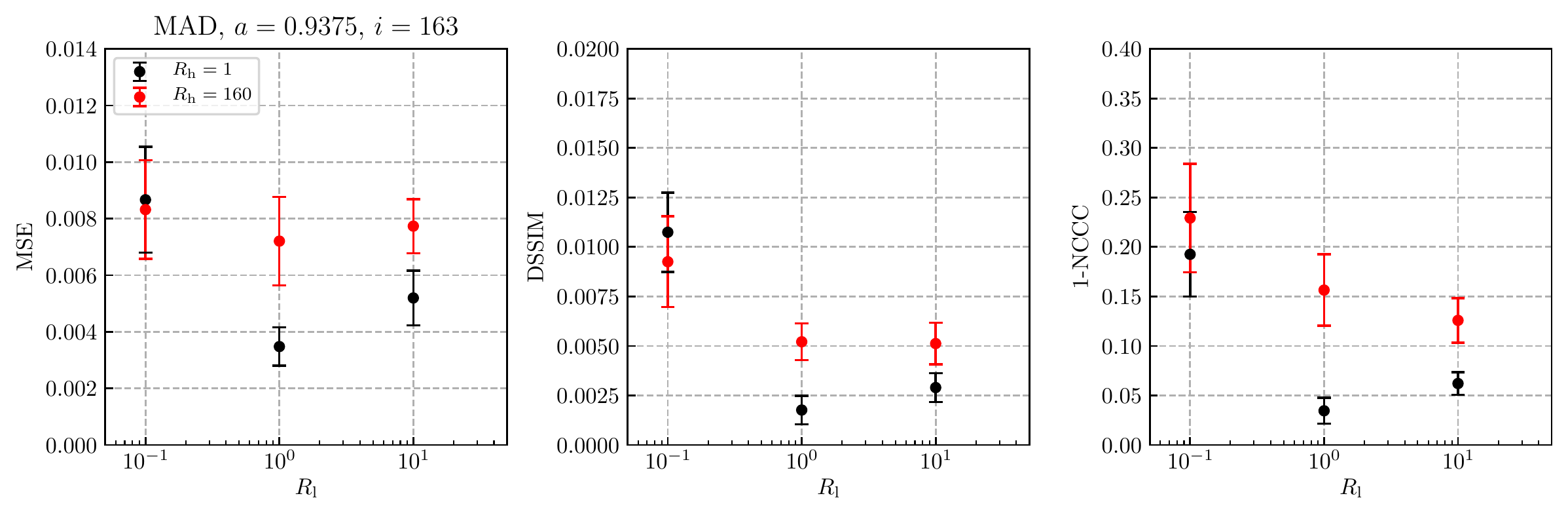}
 \caption{Image-comparison distributions between the turbulent-heating
   prescription and the parameterised ion-to-electron ratio prescription
   with different values of $R_\mathrm{l}$ and a fixed $R_\mathrm{h}=1$
   (black) and $160$ (red) for a black hole with spin
   $a=0.9375$. Different points indicate the mean value and the shading
   the standard deviation in the variation.}
 \label{fig:comp-image_i163_Rl}
\end{center}
\end{figure*}
%

The image comparison results for the 230\,GHz GRRT images at $i =
163\degr$ are shown in Fig.~\ref{fig:comp-image_i163_Rl}. Overall, when
$R_\mathrm{l}$ is large, all of the three image-comparison metrics become
worse, in particular for the $R_{\rm h}=1$ case.
This occurs since the emission structure differs significantly from the
heating-prescription case. On the other hand, if $R_\mathrm{l} \lesssim
1$, all of the three metrics increase. This could be understood as due to
the extended emission, which is even more extended than in the
electron-heating prescriptions.
 This behaviour does not change considerably when considering different
values of $R_\mathrm{h}$. Hence, from these results we conclude that our
default value $R_{\mathrm l}=1$ effectively provides a very good match
with the electron-heating prescriptions.

\section{Mass-Accretion Rate}
\label{sec:mass_acc_rate}

We recall that although the GRMHD simulations are scale-free, the GRRT
calculations depend on the physical scale set by the mass of the black
hole. In this work, our reference black hole is M87*. A mass-scale unit
is needed in the conversion from the value of the rest-mass employed in
the simulations to a physical rest-mass density; this is done by
normalising the time-averaged flux at 230~GHz to the value of
0.5\,Jy. Once a physical value for the rest-mass density is obtained, we
can calculate a physical mass-accretion rate.

\begin{figure}
\begin{center}
 \includegraphics[width=1.0\linewidth]{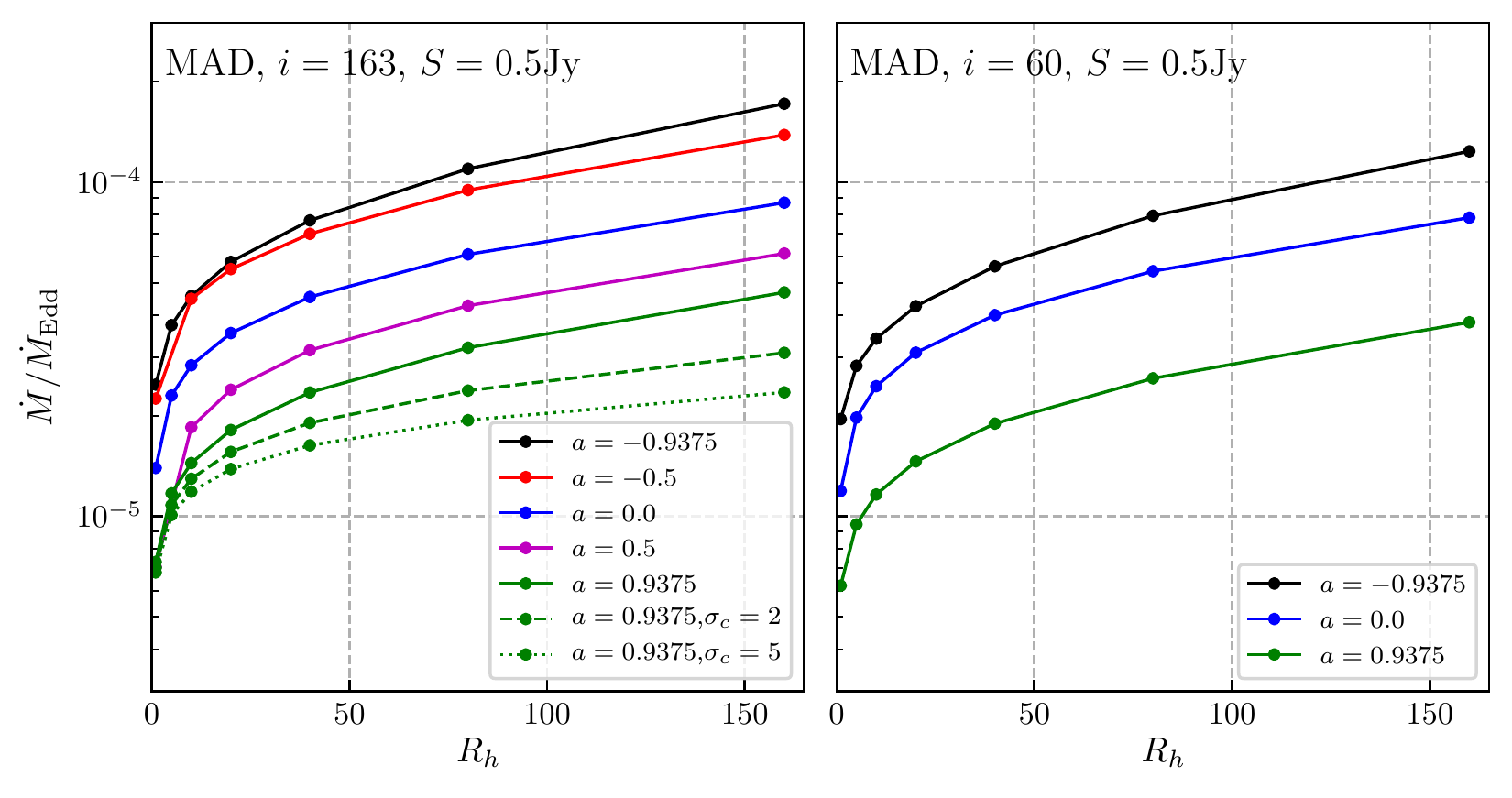}
 \caption{Mass-accretion rate normalised by the Eddington mass-accretion
   rate for different $R_\mathrm{h}$ values when considering different
   black-hole spins.  The mass-accretion rate is computed employing the
   mass and distance in M87* and by setting the time-averaged flux at
   230~GHz to the value of 0.5\,Jy.}
 \label{fig:mass_acc_rate}
\end{center}
\end{figure}

Figure~\ref{fig:mass_acc_rate} shows the mass-accretion rates normalised
by the Eddington mass-accretion rate in different models. Clearly, the
mass-accretion rate increases monotonically with the values of
$R_\mathrm{h}$ value and decreases as the black-hole spin goes from
maximally counter-rotating to maximally co-rotating; this latter
behaviour is possibly due to the increase of the ISCO in the
counter-rotating case, which allows larger-density material to be
accreted. Overall, the mass-accretion rate is in the range from $10^{-6}$
to $10^{-4}$ times Eddington accretion rate. Interestingly, higher
$\sigma$ cutoffs lead to lower mass-accretion rates; this is probably
because when using a larger $\sigma$ cutoff we are including larger
regions with high magnetisation, where the rest-mass densities are small
but contribute to reaching the same reference flux.
 %
%


\bsp	
\label{lastpage}
\end{document}